\documentclass[aps,prb,twocolumn,showpacs,groupedaddress,
longbibliography]{revtex4-1}

\usepackage[T1]{fontenc}
\usepackage[english]{babel}
\usepackage{epsfig}
\usepackage{graphicx}
\usepackage{color}
\usepackage{amsmath}
\usepackage{bbold}
\usepackage{float}
\usepackage{subfigure}

\newcommand{\bs}[1] {\mathbf{#1}}
\newcommand{\ve}{V}
\newcommand{\fl} [2]{#1^{\Lambda}_{\mathrm{#2} }  }

\begin{document}

\title{Non-separable frequency dependence of two-particle vertex \\ 
        in interacting fermion systems}

\author {Demetrio~Vilardi}
\author{Ciro~Taranto}
\author{Walter~Metzner}
\affiliation{Max Planck Institute for Solid State Research, Heisenbergstrasse 1, D-70569 Stuttgart, Germany}

\date{\today}

\begin{abstract}
We derive functional flow equations for the two-particle vertex and the self-energy in interacting fermion systems which capture the full frequency dependence of both quantities. The equations are applied to the hole-doped two-dimensional Hubbard model as a prototype system with entangled magnetic, charge and pairing fluctuations. Each fluctuation channel acquires substantial dependencies on all three Matsubara frequencies, such that the frequency dependence of the vertex cannot be accurately represented by a channel sum with only one frequency variable in each term. At the temperatures we are able to access, the leading instabilities are mostly antiferromagnetic, with an incommensurate wave vector. However, at large doping, a divergence in the charge channel occurs at a finite frequency transfer, if the vertex flow is computed without self-energy feedback. This enigmatic instability was already observed in a calculation by Husemann \emph{et al.} [Phys.~Rev.~B \textbf{85}, 075121 (2012)], who used an approximate separable ansatz for the frequency dependence of the vertex. We identify a simple mechanism for this instability in terms of a random phase approximation for the charge channel with a frequency dependent effective magnetic interaction as input.
In spite of the strong momentum and frequency dependence of the vertex, the self-energy has a Fermi liquid form.
At the moderate interaction strength where our approach is applicable, we obtain a moderate reduction of the quasi-particle weight and a sizable decay rate with a pronounced momentum dependence. Nevertheless, the self-energy feedback into the vertex flow turns out to be crucial, as it suppresses the unphysical finite frequency charge instability.
\end{abstract}

\pacs{}
\maketitle


\section{Introduction}
\label{sec:introduction}

Exact flow equations describing the evolution of correlation functions upon a successive scale-by-scale evaluation of functional integrals have become a powerful source of new approximation methods in statistical field theory \cite{Berges2002} and in the theory of quantum many-body systems -- especially interacting Fermi systems. \cite{Metzner2012} Among the various versions of these Wilsonian flows, which go under the name {\em functional renormalization group} (fRG), Wetterich's \cite{Wetterich1993} flow equation for the generating functional of one-particle irreducible vertex functions turned out to be particularly efficient. 
While (approximate) non-perturbative solutions of the flow equations are possible for interacting bosons, for fermions one has to rely on an expansion in the fields, truncating the exact hierarchy of flow equations beyond $m$-particle vertex functions of a certain order. One may, however, expand around a non-perturbative starting point, such as the dynamical mean-field solution. \cite{Taranto2014}

The two-particle vertex is a key quantity in any fermionic fRG flow, as it determines the two-particle correlations, leading instabilities, and also the flow of the self-energy. Unfortunately, in quantum systems the two-particle vertex is a difficult object to deal with, due to its dependence on three momentum and frequency arguments. In weakly interacting Fermi systems one may discard the frequency dependence and the momentum dependence perpendicular to the Fermi surface, as these are irrelevant in power counting. This simplification was the basis for early fRG studies of the two-dimensional Hubbard model, using an approximate static parametrization of the vertex, with a momentum dependence discretized by partitioning the Brillouin zone in patches. \cite{Zanchi1996,Halboth2000,Halboth2000b,Honerkamp2001}
Later alternative treatments of the momentum dependence using expansions with form factors were devised.\cite{Husemann2009,Eberlein2013,Eberlein2014}

While irrelevant in power counting, the frequency dependence of the vertex becomes important upon approaching instabilities toward symmetry breaking in the flow.\cite{Husemann2012} Even for weak bare interactions the two-particle vertex becomes large in that regime and acquires singular frequency dependences, for example those associated with the Goldstone boson.\cite{Eberlein2013}
Effective electron-electron interactions generated by phonon exchange also carry a frequency dependence which is physically relevant, since it describes the retardation of these interactions, and has been taken into account in renormalization group studies of the Holstein-Hubbard model.\cite{Honerkamp2007,Tam2007,Bakrim2010}
The frequency dependence of the vertex plays an increasingly important role at strong coupling, as has been confirmed for quantum impurity models,\cite{Kinza2013,Wentzell2016a} and in the dynamical mean field theory (DMFT).\cite{Georges1996,Rohringer2012}
Hence, a proper treatment of the frequency dependence of the vertex is mandatory for methods dealing with the interplay between fluctuations in all the channels at strong coupling, such as the combination of DMFT and fRG (DMF$^2$RG), \cite{Taranto2014} and other non-local diagrammatic extensions of the DMFT.\cite{Rohringer2017}

A simplified treatment of the frequency dependence, based on an additive decomposition of the two-particle vertex in pairing, magnetic and charge fluctuation channels, was developed by Husemann et al., \cite{Husemann2012} and applied to an fRG flow for the two-dimensional Hubbard model. Extending earlier work for the single-impurity Anderson model by Karrasch et al., \cite{Karrasch2008} they devised an approximate parametrization where the dependence of the vertex on the three fermionic frequencies is assumed to be {\em separable}, that is, each channel depends only on one bosonic transfer frequency, a linear combination of two fermionic frequencies. Already at this level the frequency dependence turned out to be important even at moderate coupling strengths, affecting significantly the energy scale of the leading instabilities. Moreover, for some model parameters an unexpected divergence without any plausible physical interpretation was found in the charge channel at zero momentum and {\em finite} frequency transfer. \cite{Husemann2012}
In a refined parametrization Husemann et al.\ also found pronounced dependences on the remaining fermionic frequencies, but with little influence on the instability scales. \cite{Husemann2012}  
At about the same time, Uebelacker and Honerkamp \cite{Uebelacker2012} evaluated the fRG flow of the frequency dependent vertex and self-energy without making any separability assumptions, albeit with a relatively rough discretization (10 points) of the Matsubara axis. For their choice of model parameters only a moderate reduction of the instability scales was observed.

In this paper we present fRG flows for the two-particle vertex and the self-energy, where the frequency dependence is fully taken into account with a high resolution and without simplifying assumptions. The two-dimensional Hubbard model is used as a prototype fermion system featuring strong and competing fluctuations in several channels. We demonstrate the feasibility, and in some respects, also the necessity of a computation with an unbiased frequency parametrization, even at moderate coupling. Significant {\em non-separable} frequency dependences appear. The various interaction channels do not depend on the bosonic transfer frequencies only, but also on the remaining two fermionic frequencies. We recover the enigmatic charge instability discovered by Husemann et al., \cite{Husemann2012} and reveal its mechanism as the impact of a frequency dependent magnetic interaction on the charge channel.

While a static vertex entails a static self-energy in the one-particle irreducible fRG formalism, the implementation of the full dynamics allows us to compute the frequency (and momentum) dependence of the self-energy. Most interestingly, the feedback of the self-energy into the flow equation for the vertex eliminates the unphysical divergence in the charge channel. This is in contrast with the widespread assumption that the self-energy feedback plays a minor role at moderate interaction strengths.

The paper is structured as follows. In Sec.~\ref{sec:formalism} we will introduce the two-dimensional Hubbard model and the fRG flow equations for the two-particle vertex and the self-energy.
After discussing the channel decomposition and our parametrization of the two-particle vertex  in Sec.~\ref{sec:vertex}, we will move on to the discussion of the main results in Sec.~\ref{sec:results}. Here we identify the leading instabilities, and we discuss the flow of the frequency-dependent vertex. For the charge divergence we provide a transparent explanation, and we finally discuss the momentum and frequency dependence of the self-energy. We draw our conclusions in Sec.~\ref{sec:conclusions}. In the Appendix \ref{sec:FlowEquations} we report detailed expressions for the vertex flow equations.


\section{Formalism}
\label{sec:formalism}

\subsection{Model}

The Hubbard model\cite{Montorsi1992} describes spin-$\frac{1}{2}$ fermions with a local interaction:
\begin{equation}
 \mathcal{H} = \sum_{i,j,\sigma} t_{ij} c^{\dagger}_{i,\sigma} c_{j,\sigma}
 + U \sum_{i} n_{i,\uparrow} n_{i,\downarrow} ,
\end{equation}
where $c^{\dagger}_{i,\sigma}$ ($c_{i,\sigma}$)  creates (annihilates) a fermion on site $i$ with spin orientation $\sigma$ ($\uparrow$ or $\downarrow$). We consider the two-dimensional case on a square lattice and repulsive interaction $U>0$ at finite temperature $T$. The hopping amplitude is restricted to $t_{ij} = -t$ for nearest neighbors, $t_{ij}=-t'$ for next-to-nearest neighbors. Fourier transforming the hopping matrix yields the bare dispersion relation
\begin{equation}
 \varepsilon_{\mathbf{k}} =
 -2t \left( \cos{k_x} + \cos{k_y} \right) -4 t' \cos{k_x} \cos{k_y} .
\end{equation}


\subsection{Flow equations}

In this paragraph we will provide some details about the functional renormalization group for interacting fermion systems,\cite{Metzner2012,Platt2013} defining in particular the notation used for the vertex. 

The fRG implements a scale-by-scale evaluation of the functional integral describing the many-body system. 
This is done by endowing the bare action with an additional dependence on a scale-parameter $\Lambda$,
\begin{equation}
 \mathcal{S}^\Lambda[\overline\psi,\psi] =
 -(\overline\psi,{G_0^\Lambda}^{-1}\psi)+\mathcal{S}_{\mathrm{int}},  
\end{equation} 
where $\mathcal{S}_{\mathrm{int}}$ is the interaction part, and $(\overline\psi,\psi)$ denotes the summation over all the quantum numbers of the fermionic fields  $\overline \psi$ and $\psi$. 
The scale dependence, acquired through the non-interacting propagator $G_0^\Lambda$, generates flow equations (with known initial conditions) for generating functionals. These are defined via functional integrals with the action $\mathcal{S}^\Lambda$. 
Examples are the generating functional for the connected Green's function and its Legendre transform, the so-called average effective action.\cite{Wetterich1993}
The final result is recovered for some final $\Lambda$-value restoring the original bare propagator, $G_0^{\Lambda_\mathrm{f}} = G_0$, so that the physical action of interest is recovered.  

We will apply this approach to the effective action, whose expansions in the fields generates the one-particle irreducible (1PI) vertex functions. By expanding the functional flow equation,\cite{Wetterich1993} one obtains a hierarchy of flow equations for the 1PI functions, involving vertices of arbitrarily  high orders. 
We will restrict ourselves to the two-particle level truncation by retaining only the two lowest nonvanishing orders in the expansion, that is, we consider the flow of the self-energy $\Sigma^\Lambda$ and of the two-particle vertex $V^\Lambda$, neglecting the effects of higher order vertices. 
This truncation restricts the applicability of the approach to the weak-to-moderate coupling regime.\cite{Salmhofer2001} 
It can be further shown that, at the two-particle level truncation, the fRG sums up efficiently, although approximately, the so-called parquet-diagrams.\cite{Kugler2017}
  
Due to SU(2) spin-rotation symmetry, the self-energy is diagonal in spin-space: 
\begin{equation}
\Sigma^\Lambda_{\sigma\sigma'}(k)=\Sigma^{\Lambda}(k)\delta_{\sigma,\sigma'}, 
\end{equation}
where $k=(\mathbf{k},\nu)$, $\nu$ is a fermionic Matsubara frequency and $\mathbf{k}$ a momentum in the first Brillouin zone. 
\begin{figure}[t!]
\includegraphics[width=0.3\textwidth]{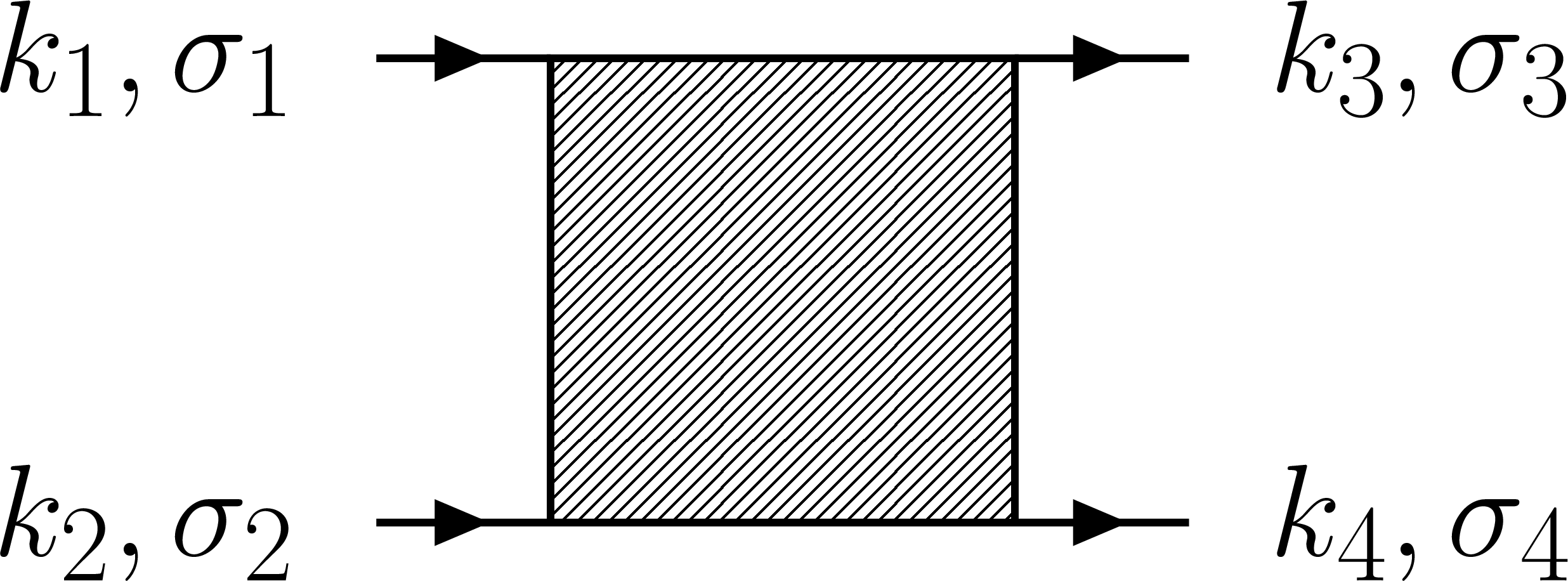}
\caption{Notation of the two-particle vertex.} 
\label{fig:notvert} 
\end{figure}

For the notation of the two-particle vertex function $V^{\Lambda}_{\sigma_1\sigma_2\sigma_3\sigma_4}(k_1,k_2,k_3)$ we refer to Fig.~\ref{fig:notvert}, where $k_i=(\mathbf{k_i},\nu_i)$.
The momentum $k_4=k_1+k_2-k_3$ is fixed by momentum conservation.
The SU(2) spin-rotation symmetry guarantees that the vertex vanishes for all spin combinations but six:
$V^\Lambda_{\uparrow\uparrow\uparrow\uparrow} =
 V^\Lambda_{\downarrow\downarrow\downarrow\downarrow}$, 
$V^\Lambda_{\uparrow\downarrow\uparrow\downarrow} =
 V^\Lambda_{\downarrow\uparrow\downarrow\uparrow}$, and
$V^\Lambda_{\uparrow\downarrow\downarrow\uparrow } =
 V^\Lambda_{\downarrow\uparrow\uparrow\downarrow}$.   
Finally, due to SU(2) symmetry and crossing relation one has \cite{Rohringer2012} 
\begin{eqnarray}
\nonumber
V^\Lambda_{\uparrow\uparrow\uparrow\uparrow}(k_1,k_2,k_3) &=& V^\Lambda_{\uparrow\downarrow\uparrow\downarrow}(k_1,k_2,k_3)\\&-& V^\Lambda_{\uparrow\downarrow\uparrow\downarrow}(k_1,k_2,k_1+k_2-k_3),
\label{eq:spinsym1}
 \\ 
V^\Lambda_{\uparrow\downarrow\downarrow\uparrow}(k_1,k_2,k_3)& =& -V^\Lambda_{\uparrow\downarrow\uparrow\downarrow}(k_1,k_2,k_1+k_2-k_3).
\label{eq:spinsym2}
\end{eqnarray}
This allows us to express the vertex by only one function of three frequency-momentum arguments: $V^\Lambda(k_1,k_2,k_3)\equiv V^\Lambda_{\uparrow\downarrow\uparrow\downarrow}(k_1,k_2,k_3)$.\cite{Husemann2009}

\begin{figure}[t!]
\includegraphics[width=0.4\textwidth]{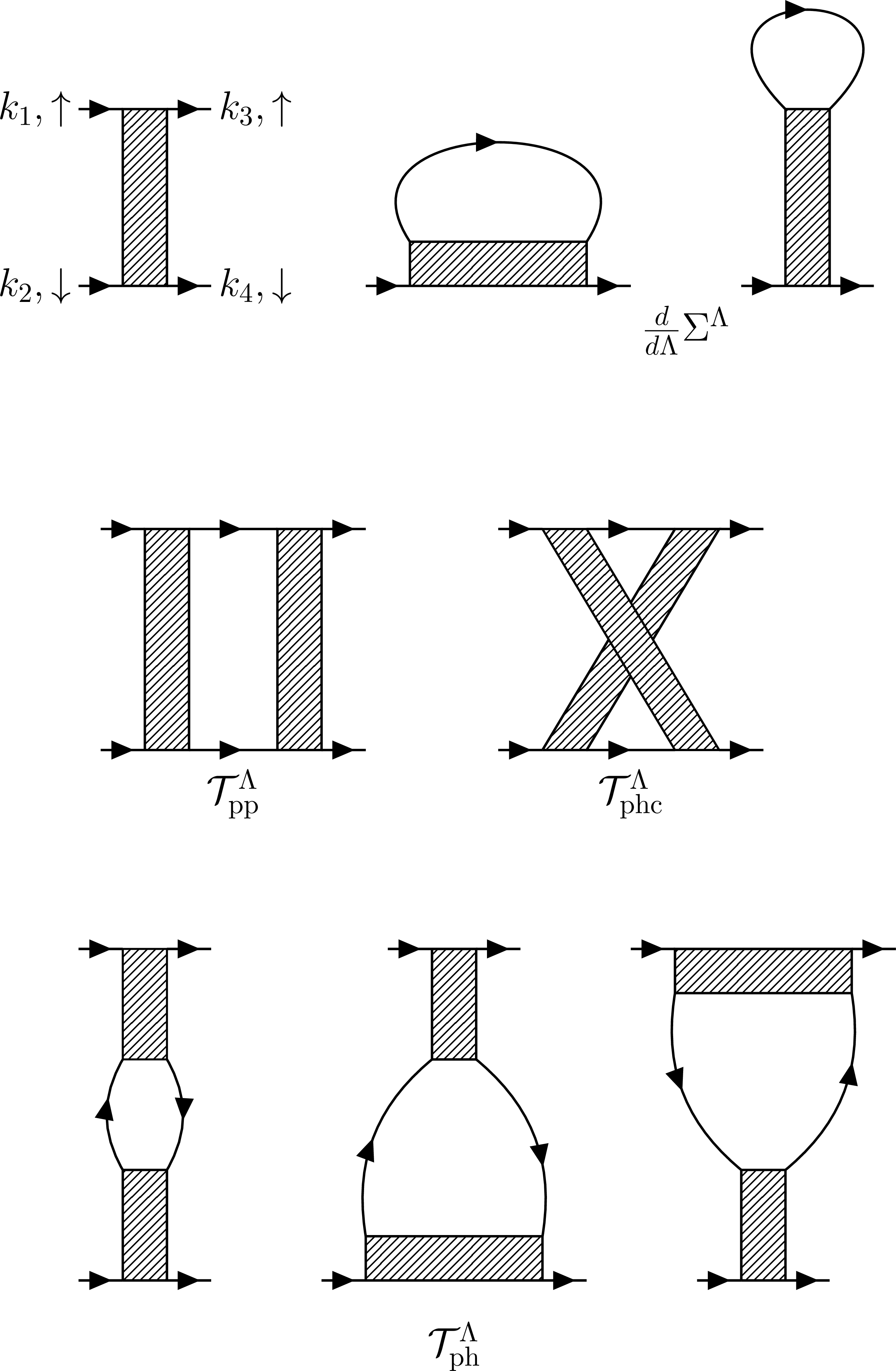}
\caption{Top row: vertex spin convention for $V^{\Lambda}(k_1,k_2,k_3)$ and diagrams contributing to the self-energy.
Second and third rows: diagrams for the flow of the vertex function. 
The internal lines are either full propagators or single-scale propagators. } 
\label{fig:diagrams} 
\end{figure}

The flow equation for the self energy can then be written as \cite{Metzner2012} 
\begin{equation}
\frac{d}{d \Lambda} \Sigma^\Lambda(k)= \int_p  S^\Lambda(p)\left[2V^\Lambda(k,p,p) -V^\Lambda(k,p,k)\right], 
\end{equation}
with $p=(\mathbf{p},\omega)$ and $k = (\mathbf{k},\nu)$.
For a diagrammatic representation, see Fig.~\ref{fig:diagrams}.
We use the notation  $\int_{p} = T \sum_\omega \int_{\mathbf{p}}$, where $\sum_\omega$ is the Matsubara frequency sum, and $\int_{\mathbf{p}}=\int  \frac{d\mathbf{p}}{(2\pi)^2}$ is the normalized integration over the first Brillouin zone. 
\begin{equation}
 S^\Lambda = \left. \frac{dG^\Lambda}{d\Lambda}\right|_{\Sigma^{\Lambda}=\mathrm{const}} 
\end{equation}
is the so-called single-scale propagator, and ${G^\Lambda}$ is the full propagator, which is related to the bare propagator and the self-energy by the Dyson equation
$(G^\Lambda)^{-1} = (G_0^\Lambda)^{-1} - \Sigma^\Lambda$. 
  
\begin{widetext} 
The flow equation for the vertex can be written as \cite{Metzner2012, Husemann2009}
\begin{align}
\label{eq:vertflow}
 \frac{d}{d\Lambda}V^\Lambda(k_1,k_2,k_3) = \fl{\mathcal{T}}{pp}(k_1,k_2,k_3) +  
  \fl{\mathcal{T}}{ph}(k_1,k_2,k_3) + \fl{\mathcal{T}}{phc}(k_1,k_2,k_3),
\end{align} 
where
\begin{eqnarray}
\label{eq:ppT} 
\fl{\mathcal{T}}{pp}(k_1,k_2,k_3) &=& - \int_p \fl{\mathcal{P}}{\mathrm{pp}}(k_1+k_2,p) \fl{V}{}(k_1,k_2,k_1+k_2-p)\fl{V}{}(k_1+k_2-p,p,k_3) ,
\label{eq:tpp} 
\\ 
\label{eq:tph} 
\fl{\mathcal{T} } {ph}(k_1,k_2,k_3) & =& \int_p \fl{\mathcal{P}}{\mathrm{ph}}(k_3-k_1,p)
\Big\{ 2 \fl{V}{}( k_1,k_3-k_1+p,k_3)  \fl{V}{}(p,k_2,k_3-k_1+p) \\
\nonumber
&&- \fl{V}{}( k_1,k_3-k_1+p,p)  \fl{V}{}(p,k_2,k_3-k_1+p) - \fl{V}{}( k_1,k_3-k_1+p,k_3)  \fl{V}{}(k_2,p,k_3-k_1+p) \Big\} , \\
\label{eq:tphc}
\fl{\mathcal{T}}{phc}(k_1,k_2,k_3) & =& -\int_p \fl{\mathcal{P}}{\mathrm{ph}}(k_2-k_3,p) \fl{V}{}(k_1,k_2-k_3+p,p)
\fl{V}{}(p,k_2,k_3).
\end{eqnarray} 
\end{widetext}
Here $\mathcal{T}^\Lambda_{\mathrm{pp}}$, $\mathcal{T}^\Lambda_{\mathrm{ph}}$ and $\mathcal{T}^\Lambda_{\mathrm{phc}}$ stand respectively for \textit{particle-particle}, \textit{particle-hole} and \textit{particle-hole crossed} contributions.
We have defined the quantities
\begin{align}
 \mathcal{P}_{\mathrm{ph}}^\Lambda(Q,p) &= G^\Lambda(Q+p)S^\Lambda(p) + G^\Lambda(p) S^\Lambda(Q+p), \\ 
 \mathcal{P}_{\mathrm{pp}}^\Lambda(Q,p) &=G^\Lambda(Q-p)S^\Lambda(p) + G^\Lambda(p) S^\Lambda(Q-p) ,
\end{align} 
which are the scale derivatives, at fixed self-energy, of the product of two Green's functions. The diagrams contributing to the vertex flow are represented in Fig.~\ref{fig:diagrams}.


\subsection{Interaction flow}
\label{sec:IntFlow}
To use the flow equations defined above we need to specify the $\Lambda$-dependence of the non-interacting propagator $G_0^\Lambda$.
We use the \textit{interaction flow}, introduced by Honerkamp {\em et al.}: \cite{Honerkamp2004}
 \begin{equation}
 G_0^\Lambda(k) = \Lambda G_0(k) =
 \frac{\Lambda}{i\nu + \mu^\Lambda - \varepsilon_{\mathbf{k}}} , 
 \end{equation}
where the scale-parameter $\Lambda$ flows from $0$ to $1$. 
We have introduced a $\Lambda$-dependent chemical potential to maintain the density fixed during the flow.
The Dyson equation yields the interacting Green's function in the form
\begin{equation}
 G^\Lambda(k) = \frac{\Lambda}{i\nu - \varepsilon_\bs{k}
 + \mu^{\Lambda} - \Lambda\Sigma^\Lambda(k)} . 
\end{equation}
The corresponding single-scale propagator is given by
\begin{equation}
 S^\Lambda(k) = 
 \frac{i\nu - \varepsilon_\bs{k} + \mu^\Lambda - \Lambda \partial\mu^\Lambda/\partial\Lambda}
 {\left(i\nu - \varepsilon_\bs{k} + \mu^\Lambda - \Lambda \Sigma^\Lambda \right)^2} .
\end{equation}
The scale-dependent chemical potential $\mu^\Lambda$ is determined from the equation
\footnote[4]{Note that in the interaction flow $G^\Lambda$ is defined  with fermion fields
which are rescaled by a factor $\sqrt{\Lambda}$. \cite{Honerkamp2004} To compute physical quantities
from $G^\Lambda$ one therefore has to undo this rescaling and divide the
propagator by $\Lambda$. }
\begin{equation}
n = n^{\Lambda}(\mu^\Lambda) \equiv 2 \int_k \frac{e^{i\nu 0^{+}}} {i\nu - \varepsilon_{\bs{k}} +\mu^{\Lambda}-\Lambda\Sigma^\Lambda(k)}. 
\end{equation} 
The factor 2 accounts for the spin degree of freedom.
Without self-energy feedback in the propagator, $\mu^\Lambda$ is constant. The scale dependence of $\mu^\Lambda$ generated by the self-energy is rather weak for the model parameters chosen in our calculations, so that we could neglect the term proportional to $\partial\mu^\Lambda/\partial\Lambda$ in the single-scale propagator.

The main advantage of the interaction flow is that the $\Lambda$-dependent action can be interpreted \cite{Honerkamp2004} as the physical action of the system with a rescaled interaction $U^\Lambda = \Lambda^2 U$.
Since $T$ acts as an infrared cutoff, for our purposes we do not need to worry about the fact that this flow is not scale-selective, and hence is not regularizing infrared divergences. 


\section{Vertex parametrization}
\label{sec:vertex}

To parametrize the momentum and frequency dependence of the two-particle vertex, we use the channel decomposition of the vertex introduced by Husemann and Salmhofer, \cite{Husemann2009} where the vertex is written as a sum of the bare interaction and fluctuation induced effective pairing, magnetic and charge interactions.
The function $V^\Lambda(k_1,k_2,k_3)$ is thus decomposed as
\begin{eqnarray}
\nonumber
\ve^{\Lambda}(k_1,k_2,k_3)&=& U - \phi^{\Lambda}_{\mathrm{p}}(k_1+k_2;k_1,k_3)  \\
&+& \phi^{\Lambda}_{\mathrm{m}}(k_2-k_3;k_1,k_2)  \nonumber
 \\ 
 &+&
  \frac{1}{2}  \phi^{\Lambda}_{\mathrm{m}}(k_3-k_1;k_1,k_2) \nonumber \\ & -& \frac{1}{2} \phi^{\Lambda}_{\mathrm{c}}(k_3-k_1;k_1,k_2),
 \label{eq:decomposition}
\end{eqnarray}
in terms of the {\em pairing} channel $\phi_{\mathrm{p}}$, the {\em magnetic} channel $\phi_{\mathrm{m}}$ and the {\em charge} channel $\phi_{\mathrm{c}}$. The first argument of $\phi_{\mathrm{p}}$ is the conserved total momentum and frequency of the particles, while the first argument of $\phi_{\mathrm{m}}$ and $\phi_{\mathrm{c}}$ is a momentum and frequency transfer.
Substituting Eq.~(\ref{eq:decomposition}) into Eq.~(\ref{eq:vertflow}) we obtain: 
\begin{eqnarray}
\nonumber
&&-\dot \phi^{\Lambda}_{\mathrm{p}}(k_1+k_2;k_1,k_3)+ \dot \phi^{\Lambda}_{\mathrm{m}}(k_2-k_3;k_1,k_2)\\&& 
 + \frac{1}{2}  \dot\phi^{\Lambda}_{\mathrm{m}}(k_3-k_1;k_1,k_2)- \frac{1}{2} \dot\phi^{\Lambda}_{\mathrm{c}}(k_3-k_1;k_1,k_2)\\&&=
   \fl{\mathcal{T}}{pp}(k_1,k_2,k_3) +  
  \fl{\mathcal{T}}{ph}(k_1,k_2,k_3) + 
  \fl{\mathcal{T}}{phc}(k_1,k_2,k_3).
  \nonumber
\label{eq:decot}
\end{eqnarray} 
We associate the total momentum argument of $\mathcal{P}^\Lambda_{\rm pp}$ and the momentum transfer argument of $\mathcal{P}^\Lambda_{\rm ph}$ in Eqs.~(\ref{eq:tpp}-\ref{eq:tphc}) to the corresponding arguments of the $\phi_{\mathrm{x}}$ on the right hand side of Eq.~\ref{eq:decomposition}.
This way, it is easy to attribute $\fl{\mathcal{T}}{pp}$ to the flow equation of the only function in Eq. (\ref{eq:decot}) that depends explicitly on $k_1+k_2$: $-\dot\phi_{\mathrm{p}}^\Lambda=\fl{\mathcal{T}}{pp}$. 
The same is true for the particle-hole crossed channel: $\fl{\mathcal{T}}{phc}=\dot\phi_{\mathrm{m}}^\Lambda$.  We associate to the particle-hole diagram the third and fourth term on the left hand side of Eq.~(\ref{eq:decot}): $\fl{\mathcal{T}}{ph}(k_1,k_2,k_3)=\frac{1}{2}  \dot\phi^{\Lambda}_{\mathrm{m}}(k_3- k_1;k_1,k_2) - \frac{1}{2} \dot\phi^{\Lambda}_{\mathrm{c}}(k_3-k_1;k_1,k_2) $.  
The flow equations for $\phi_{\mathrm{x}}$ then read: \cite{Husemann2009}
\begin{eqnarray}
\label{eq:phi_p}
 \dot{\phi}_{\mathrm{p}}^{\Lambda}(Q;k_1,k_3) &=&
 -\mathcal{T}^{\Lambda}_{\mathrm{pp}}(k_1,Q-k_1,k_3) , \\
\label{eq:phi_c}
 \dot{\phi}_{\mathrm{c}}^{\Lambda}(Q;k_1,k_2) &=&
 \mathcal{T}^{\Lambda}_{\mathrm{phc}}(k_1,k_2,k_2-Q) \nonumber \\
 && -2\mathcal{T}^{\Lambda}_{\mathrm{ph}}(k_1,k_2,Q+k_1), \\
\label{eq:phi_m}
\dot{\phi}_{\mathrm{m}}^{\Lambda}(Q;k_1,k_2) &=& \mathcal{T}^{\Lambda}_{\mathrm{phc}}(k_1,k_2,k_2-Q) .
\end{eqnarray}
 
Following Refs.~\onlinecite{Husemann2009,Husemann2012}, we address first the momentum dependence. To parametrize the dependence on the fermionic momenta, we use a decomposition of unity by means of a set of orthonormal form factors
$\{f_{l}(\bs{k})\}$.
We can then project each channel on a subset of form factors, whose choice is physically motivated.\cite{Husemann2009}

For the pairing channel we keep only $f_{s}(\bs{k}) = 1$ and $f_d(\bs{k})=\cos{k_x} - \cos{k_y}$:
\begin{align}
 \phi^{\Lambda}_{\mathrm{p}}(Q;k_1,k_3) &=
 \mathcal{S}^\Lambda_{\bs{Q},\Omega}(\nu_1,\nu_3) \nonumber \\ 
 &+ f_d\left(\frac{\bs{Q}}{2}-\bs{k}_1\right) f_d\left(\frac{\bs{Q}}{2}-\bs{k}_3\right) \mathcal{D}^\Lambda_{\bs{Q},\Omega}(\nu_1,\nu_3).
\end{align}
A divergence in the channel $\mathcal{S}$ ($\mathcal{D}$) is associated to the emergence of $s$-wave ($d$-wave) superconductivity.\cite{Metzner2012,Platt2013}

For the charge and magnetic channels we restrict ourselves to $f_{s}(\bs{k})=1$ only:
\begin{align}
  \phi^\Lambda_{\mathrm{c}}(Q;k_1,k_2) &= \mathcal{C}^\Lambda_{\bs{Q},\Omega}(\nu_1,\nu_2), \\
  \phi^\Lambda_{\mathrm{m}}(Q;k_1,k_2) &= \mathcal{M}^\Lambda_{\bs{Q},\Omega}(\nu_1,\nu_2).
\end{align}
A divergence of these functions signals $s$-wave instabilities in the charge and magnetic channels, respectively.

Each channel in Eq.~(\ref{eq:decomposition}) contains a (bosonic) linear combination of momenta and frequencies, and two remaining independent fermionic momentum and frequency variables. 
The choice of the mixed notation is natural since the bosonic momenta and 
frequencies play a special role in the diagrammatics.
Indeed, it is the only dependence generated in second order perturbation theory and the main dependence in finite order perturbation theory.
Although one expects a dominant dependence on the bosonic frequency, in particular in the weak coupling regime, we will see that the dependence on the fermionic frequencies can become strong and non-negligible, too.
In Refs.~\onlinecite{Husemann2009,Husemann2012}, without any or with a simplified frequency dependence, the channel functions are interpreted as bosonic exchange propagators. Such an interpretation is not possible with full frequency-dependence.

The flow equations for the channels $\mathcal{S}$,  $\mathcal{D}$, $\mathcal{C}$ and $\mathcal{M}$ can be derived from the projection of Eqs.~(\ref{eq:phi_p})-(\ref{eq:phi_m}) onto the form factors:
\begin{eqnarray}
\dot{\mathcal{S}}_{\bs{Q},\Omega}^{\Lambda}(\nu_1,\nu_3)  &=& - \int _{\bs{k}_1, \bs{k}_3 } \mathcal{T}^\Lambda_{\mathrm{pp}}(k_1,Q-k_1,k_3), \\ 
\dot{\mathcal{D}}_{\bs{Q},\Omega}^{\Lambda}(\nu_1,\nu_3)  &=& -
\int _{\bs{k}_1,\bs{k}_3}  f_d\left( {\frac{\bs{Q}}{2} - \bs{k}_1} \right ) f_d\left ({\frac{\bs{Q}}{2} - \bs{k}_3} \right)  \nonumber \\ 
 && \times \, \mathcal{T}^\Lambda_{\mathrm{pp}}(k_1,Q-k_1,k_3) , \\
\nonumber
\dot{\mathcal{C}}_{\bs{Q},\Omega}^{\Lambda}(\nu_1,\nu_2) &=& 
\int _{\bs{k}_1,\bs{k}_2}  \Big[ \mathcal{T}^\Lambda_{\mathrm{phc}}(k_1,k_2,k_2-Q)  \\
 && - \, 2\mathcal{T}_{\mathrm{ph}}(k_1,k_2,Q+k_1) \Big] , \\ 
\dot{\mathcal{M}}_{\bs{Q},\Omega}^{\Lambda}(\nu_1,\nu_2) & =& 
\int _{\bs{k}_1,\bs{k}_2}  \mathcal{T}^\Lambda_{\mathrm{phc}}(k_1,k_2,k_2-Q) . 
\end{eqnarray}
The final equations are then obtained by substituting the decomposition (\ref{eq:decomposition}) into the equations above, and using trigonometric identities.

As an example we report here the equations for the magnetic channel, while the expressions for the other channels are presented in the Appendix \ref{sec:FlowEquations}:
\begin{widetext}
\begin{equation}
 \dot{\mathcal{M}}^{\Lambda}_{\bs{Q},\Omega}(\nu_1,\nu_2) = 
 -T \sum_\nu L^{\mathrm{m},\Lambda}_{\mathbf{Q},\Omega}(\nu_1,\nu) 
 P_{\bs{Q},\Omega}^{\Lambda}(\nu) 
 L^{\mathrm{m},\Lambda}_{\mathbf{Q},\Omega}(\nu,\nu_2-\Omega), 
\label{eq:FlowMag}
\end{equation} 	   
with
\begin{equation}
 P_{\bs{Q},\Omega}^{\Lambda}(\omega) = \int_{\bs{p}}
 G^\Lambda(\bs{p},\omega) S^\Lambda(\bs{Q}+\bs{p},\Omega+\omega) +
 G^\Lambda(\bs{Q}+\bs{p},\Omega+\omega) S^\Lambda(\bs{p},\omega),
\label{eq:Pph} 
\end{equation} 
and
\begin{eqnarray} 
\nonumber
 L^{\mathrm{m},\Lambda}_{\bs{Q},\Omega}(\nu_1,\nu_2)
 &=& U + \mathcal{M}^\Lambda_{\bs{Q},\Omega}(\nu_1,\nu_2) 
 + \int_{\bs{p}} \Big \{- \mathcal{S}^\Lambda_{\bs{p},\nu_1+\nu_2}(\nu_1,\nu_1+\Omega)  
 -\frac{1}{2} \mathcal{D}^\Lambda_{\bs{p},\nu_1+\nu_2}(\nu_1,\nu_1+\Omega)
 [\cos(Q_x)+\cos(Q_y)] \\
 &+&\frac{1}{2} \Big[  \mathcal{M}^\Lambda_{\bs{p},\nu_2-\nu_1-\Omega}( \nu_1,\nu_2) 
 - \mathcal{C}^\Lambda_{\bs{p},\nu_2-\nu_1-\Omega}(\nu_1,\nu_2) \Big] 
 \Big \} .
\label{eq:Lxph} 
\end{eqnarray}
\end{widetext}
Note that after the momentum integrals in $P$ and $L$ are performed, the right hand side of Eq.~(\ref{eq:FlowMag}) can be expressed as a matrix multiplication in frequency space, where $\Omega$ and $\mathbf{Q}$ appear as parameters.

After this decomposition, the evaluation of the vertex-flow equation, depending on six arguments, is reduced to the flow of the four functions $\mathcal{S}$, $\mathcal{D}$, $\mathcal{C}$, $\mathcal{M}$, each of them depending on three frequencies and one momentum only. In order to compute these equations numerically we discretize the momentum dependence on patches covering the Brillouin zone and truncate the frequency dependence to some maximal frequency value.

We conclude this section by comparing our parametrization to other fRG schemes with frequency dependent vertices. A frequency dependent vertex was first taken into account for the single-impurity Anderson model, where the parametrization is simplified significantly by the absence of momentum variables. Hence, a straightforward parametrization with three independent fermionic Matsubara frequencies is affordable, \cite{Hedden2004} but a numerically less demanding channel decomposition with only one bosonic frequency per channel \cite{Karrasch2008} has also been employed.
The full frequency dependence could also be treated for one-dimensional chain and ladder models, since the momentum dependence was parametrized by very few variables in the spirit of the familiar one-dimensional $g$-ology.\cite{Tam2007} In another work on a one-dimensional model with a retarded phonon-mediated electron-electron interaction, the frequency integrals were approximately decoupled by taking frequency averages of coupling functions.\cite{Bakrim2010} In all these works the frequency resolution was much rougher than in our calculations.
In two dimensions, fRG calculations with three independent frequency variables were first performed by Honerkamp et al.,\cite{Honerkamp2007} and then by Uebelacker and Honerkamp,\cite{Uebelacker2012} where in the second work also self-energy feedback was taken into account. The frequency dependence was not restricted by an ansatz, but its discretization was limited to only 10 points. The momentum dependence was discretized via a large number of patches for all three fermionic momenta.
Complementary to this direct discretization of fermionic frequencies and momenta, Husemann et al.\cite{Husemann2012} parametrized the frequency and momentum dependence by decomposing the vertex in magnetic, charge and pairing channels, and approximating the frequency dependence of each channel by just one bosonic variable, as discussed already above. The momentum dependence was parametrized as in our work, with one bosonic variable per channel and few form factors for the remaining dependences. In a refined parametrization, the dependence on the remaining fermionic frequencies was partially taken into account via frequency dependent fermion-boson vertices.


\section{Results}
\label{sec:results}

In this section we present our results obtained by the fully frequency dependent fRG. All results are presented in units of the nearest-neighbor hopping \mbox{$t=1$}. Unless specified otherwise, the next-nearest-neighbor hopping is $t'=-0.32t$, the bare interaction strength $U=4t$, and the temperature $T=0.08t$.

We have implemented numerically the flow equations presented in Eqs.~(\ref{eq:FlowMag})-(\ref{eq:Lxph}) and in the Appendix. 
To take into account the distinct momentum dependences of the self-energy and the vertex, we have defined two different patching schemes of the respective Brillouin zones. 
Similarly to what is done in Ref.~\onlinecite{Husemann2009}, the vertex patching describes more accurately the corners around $(0,0)$ and $(\pi,\pi)$, where we expect the instability vectors.
For the self-energy, the most relevant physics happens in the vicinity of the Fermi surface.
Therefore we concentrate the patches along the Fermi surface and in its immediate vicinity (see Figs.~\ref{fig:occ975} and \ref{fig:occ600}), with more points close to the \textit{antinodal} region near $(\pi,0)$, relevant for antiferromagnetism.
In the calculations presented in the following we have used $29$ patches for the vertex and $44$ for the self-energy.

For the implementation of the frequency dependence we found it convenient to rewrite $\mathcal{S}$, $\mathcal{D}$, $\mathcal{C}$ and $\mathcal{M}$ as functions of three bosonic frequencies. Note that the discrete map between triples of fermionic Matsubara frequencies and bosonic Matsubara frequencies defined by linear combinations of the former is not one-to-one, that is, only a subset of possible bosonic triples represents fermionic triples. 
For each frequency argument we restricted ourselves to at least $40$ positive and $40$ negative Matsubara frequencies. Beyond these frequencies we have used the extrapolated asymptotic behavior of the coupling functions. 


\subsection{Analysis of instabilities}

By means of the fRG one can perform an instability analysis of the system: 
for some value of the flow parameter $\Lambda$ one of the channels diverges. 
We refer to the value $\Lambda_{\mathrm{c}} $ at which this happens as the \textit{critical scale}. In the interaction flow, the critical scale corresponds to a critical interaction $U^{\Lambda_c} = \Lambda_c^2 U$, see Sec.~\ref{sec:IntFlow}.
From the diverging channel one can infer the leading instability of the system. 
\begin{figure}
\includegraphics[width=0.5\textwidth]{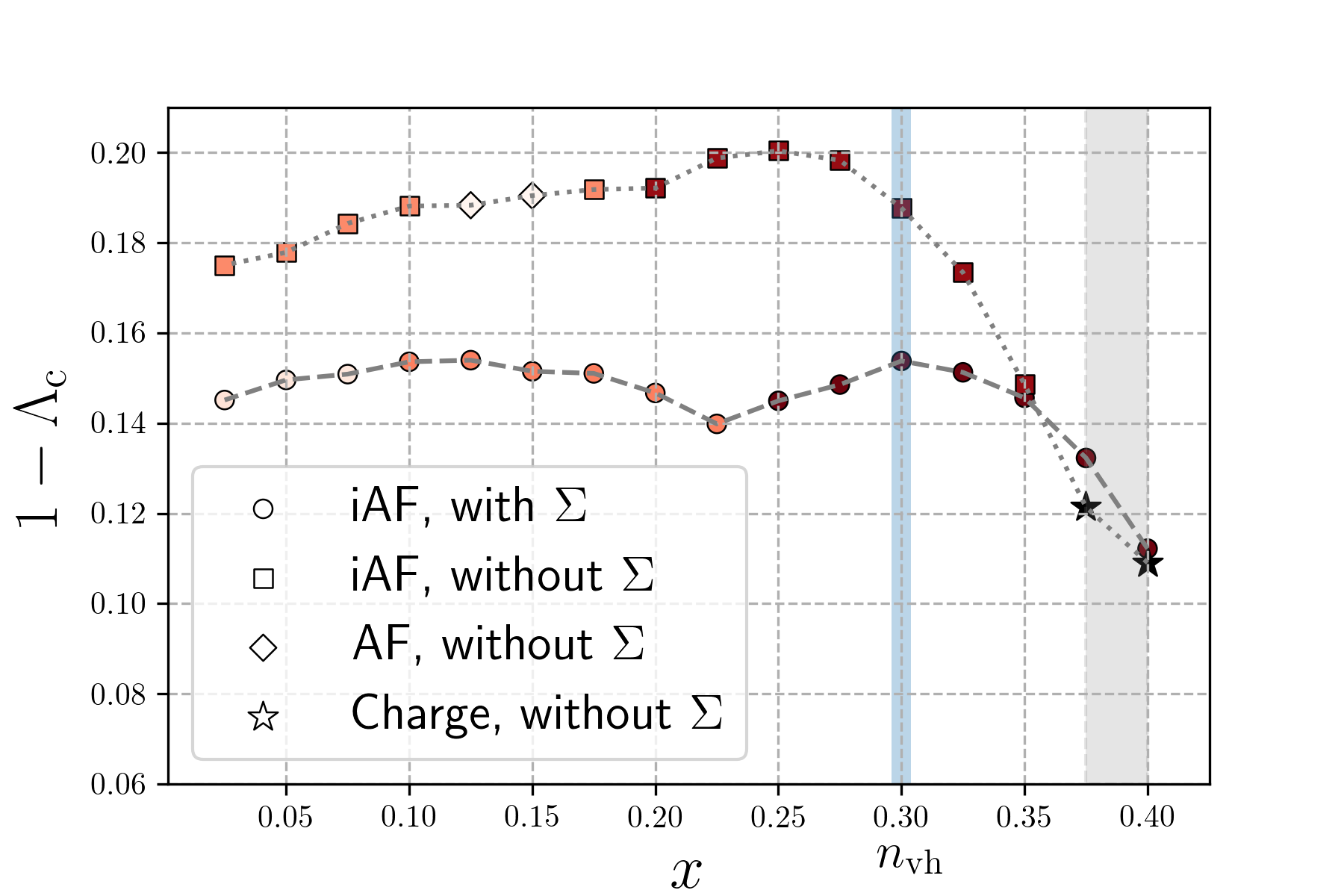}
\caption{Critical scale as a function of doping $x=1-n$, for $T = 0.08t$, $t'=-0.32t$ and $U=4t$. 
Square symbols and circles refer to incommensurate antiferromagnetism (iAF) without and with self-energy feedback, respectively.
The black stars refer to a divergence in the charge channel at $\mathbf{Q} = (0,0)$.
The color of squares and circles encodes the distance of the incommensurate magnetic $\mathbf{Q}$-vector from $(\pi,\pi)$: darker color corresponds to a larger distance. The darkest color corresponds to $\delta=1.13$. 
The vertical light blue line marks van Hove filling.}  
\label{fig:criscale} 
\end{figure}
In Fig.~\ref{fig:criscale} we show the critical scale $\Lambda_{\mathrm{c}}$ as a function of the doping $x=1-n$ with and without self-energy feedback.
We defined the critical scale as the flow parameter for which the value of the largest channel exceeds $200t$. We checked that these results are also consistent with a stability analysis based on the susceptibilities.        

A divergence of the vertex at finite temperature is associated with spontaneous symmetry breaking, in violation of the Mermin-Wagner theorem.\cite{Mermin1966}
This is a consequence of the truncation of the flow equations.
Instead, we should interpret the finite temperature vertex divergence as the signal of the appearance of strong bosonic fluctuations that cannot be treated within the approximation-scheme we are using.\cite{Salmhofer2001} 
Even though in our framework the flow cannot be continued beyond the critical scale, from the analysis of vertex and self-energy we can identify the relevant effective interactions of the system.

For the parameter sets shown in Fig.~\ref{fig:criscale}, and without self-energy feedback, there are two possible instabilities. 
For doping smaller than $0.35$ the leading fluctuations of the system are antiferromagnetic, with a commensurate (AF) wave vector $\mathbf{Q} = (\pi,\pi)$, or an incommensurate (iAF) wave vector of the form $\mathbf{Q}=(\pi,\pi-\delta)$.
The incommensurability $\delta$ is determined by the momentum $\mathbf{Q}$ where the magnetic channel $\mathcal{M}^\Lambda$ has its maximum. 
The region of commensurate antiferromagnetism for $0.125\le x \le 0.150$ has to be attributed to the presence of a large plateau around $(\pi,\pi)$ in the bare bubble. Correspondingly, the commensurate antiferromagnetic instability is almost degenerate with an incommensurate one.
\begin{figure}
\includegraphics[width=0.50\textwidth]{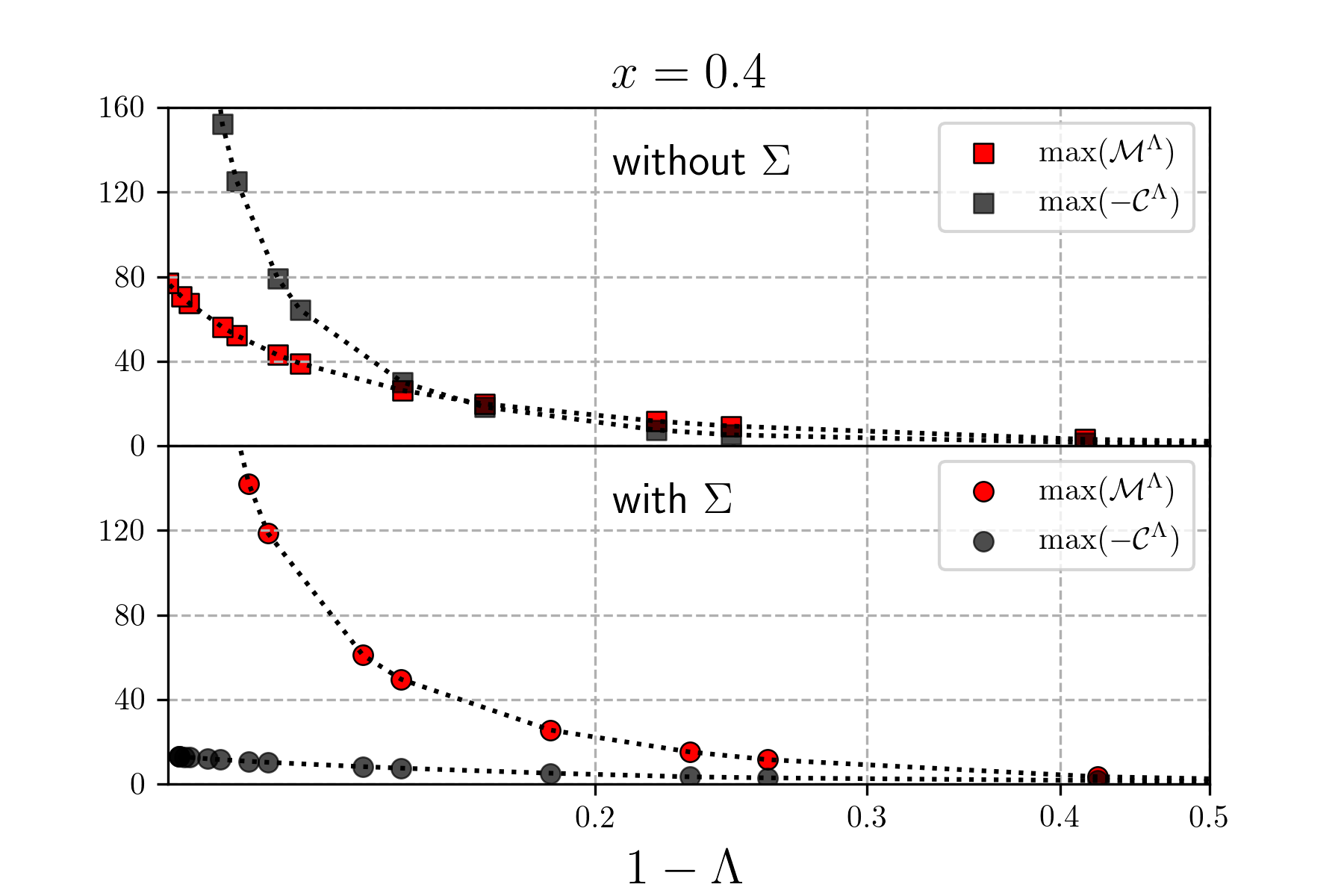}
\caption{Flow of the maximal values of the charge ($\mathcal{C}$) and magnetic ($\mathcal{M}$) channels as functions of $1-\Lambda$, for  $x=0.4$, $t'=-0.32$, $U=4t$ and $T=0.08t$.  Top: without self-energy feedback; bottom: with self-energy feedback. }
\label{fig:chargeproblem}
\end{figure}

The most striking feature in Fig.~\ref{fig:criscale} is the presence of a divergence in the charge channel $\mathcal{C}^\Lambda$ at $\mathbf{Q}=(0,0)$ for the largest values of doping, marked by black stars. 
This feature was already observed in a fRG calculation with a simplified frequency parametrization by Husemann \textit{et al.}\ in Ref.~\onlinecite{Husemann2012} and named \textit{scattering instability}. 
The charge channel $\mathcal{C}^\Lambda$ diverges for a non-zero frequency transfer $\Omega=2\pi T$, which does not allow for a natural interpretation in terms of a physical instability. 
The frequency structure of the charge channel $\mathcal{C}^\Lambda$ together with its origin will be further discussed in paragraph \ref{sec:PerpLadder}.

\begin{figure}
\includegraphics[width=0.50\textwidth]{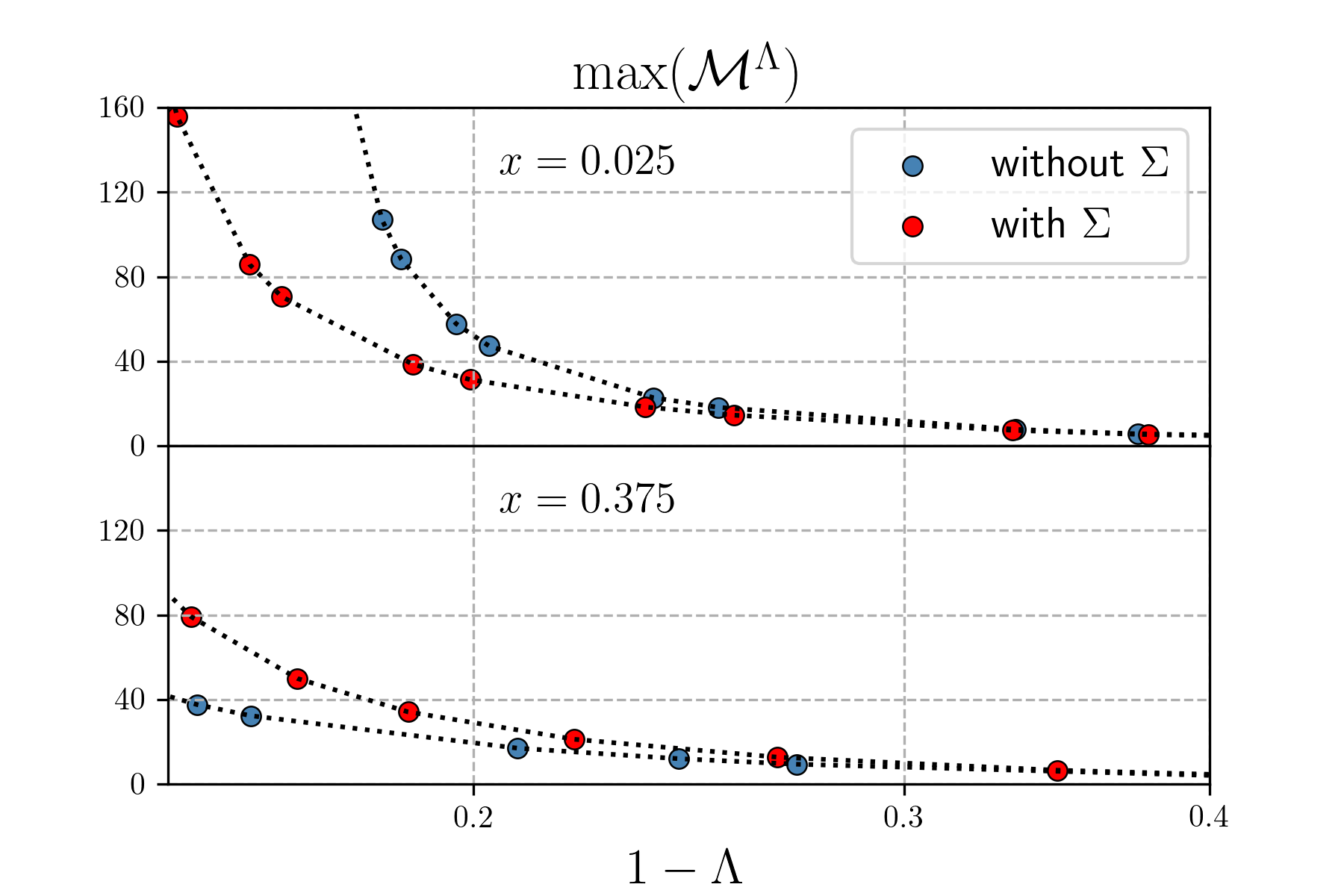}
\caption{Flow of the maximal values of the  magnetic ($\mathcal{M}$) channel as functions of $1-\Lambda$, for  $x=0.025$ (top) and $x=0.375$ (bottom). The other parameters are $t'=-0.32$, $U=4t$ and $T=0.08t$. Red symbols: with self-energy feedback; blue symbols: without self-energy feedback. }
\label{fig:selfeffect}
\end{figure}

We did not find a pairing instability at any doping. While $d$-wave pairing has been persistently obtained in most earlier fRG studies of the two-dimensional Hubbard model at sufficiently strong doping, \cite{Metzner2012} the $d$-wave pairing interaction in our calculation remains rather small. 

We attribute this seeming discrepancy to three reasons.
First, we chose a relatively high temperature to be able to accurately parametrize the frequency dependence, while the pairing interaction is expected to increase substantially only for temperatures close to the pairing scale.\footnote[3]{See, for example, the flow of the $d$-wave pairing function shown in Ref.~\onlinecite{Eberlein2014}.}
Second, as already observed by Husemann et al.,\cite{Husemann2012} previous fRG calculations with a static vertex overestimate the $d$-wave pairing channel, since the contributing effective interactions decay at large frequencies. Hence, taking the frequency dependence of the vertex into account one obtains a lower pairing scale.
Finally, in the interaction flow contributions to pairing which are discarded by our truncation at the two-particle level are more important than in the more commonly used flows with a momentum or frequency cutoff. Unlike magnetism, $d$-wave pairing is generated exclusively by diagrams with (at least two) overlapping loops. Topologically equivalent contributions can be generated from different levels of the fRG hierarchy, the only difference being the position of the single-scale propagator in the diagram. In the interaction flow these contributions have equal weight, while in cutoff flows the contributions captured already in a two-particle truncation are typically larger than those appearing only at the three-particle level.

The self-energy feedback has three effects. First, it increases $\Lambda_\mathrm{c}$, that is, it suppresses the instabilities.
Second, the incommensurability vector is affected, the region of commensurate antiferromagnetism disappears, and one can observe a more regular trend of increasing $\delta$ with $x$.
Third, the divergence in the charge channel is completely suppressed, and the leading instability in the doping region $0.375\le x \le 0.4$ remains incommensurate antiferromagnetism. 
This can be also seen from Fig.~\ref{fig:chargeproblem}, where we compare the flow of the maximum (of the absolute value) of magnetic and charge channels with and without the self-energy feedback for doping $x=0.4$.
Without self-energy feedback, the charge channel reaches large and negative values.
The presence of such a large (and negative) charge channel inhibits the magnetic channel.    
The effect of the self-energy in the flow is evident: the charge channel is strongly damped.  
At the same time the magnetic channel is enhanced.

This is confirmed by Fig.~\ref{fig:selfeffect}, where we show the maximum of $\mathcal{M}$ with and without self-energy feedback for $x=0.025$ (top) and $x=0.375$ (bottom).  
One can see that the enhancement of $\mathcal{M}$ due to the self-energy is specific of the large doping region, while, in the small doping region the self energy decreases $\mathcal{M}$.  
The self-energy affects the magnetic channel directly by reducing the particle-hole bubble, and indirectly through the feedback of other channels, that is, reducing the charge channel. The former effect dominates for small doping, the latter at large doping.

\begin{figure}
\includegraphics[width=0.5\textwidth]{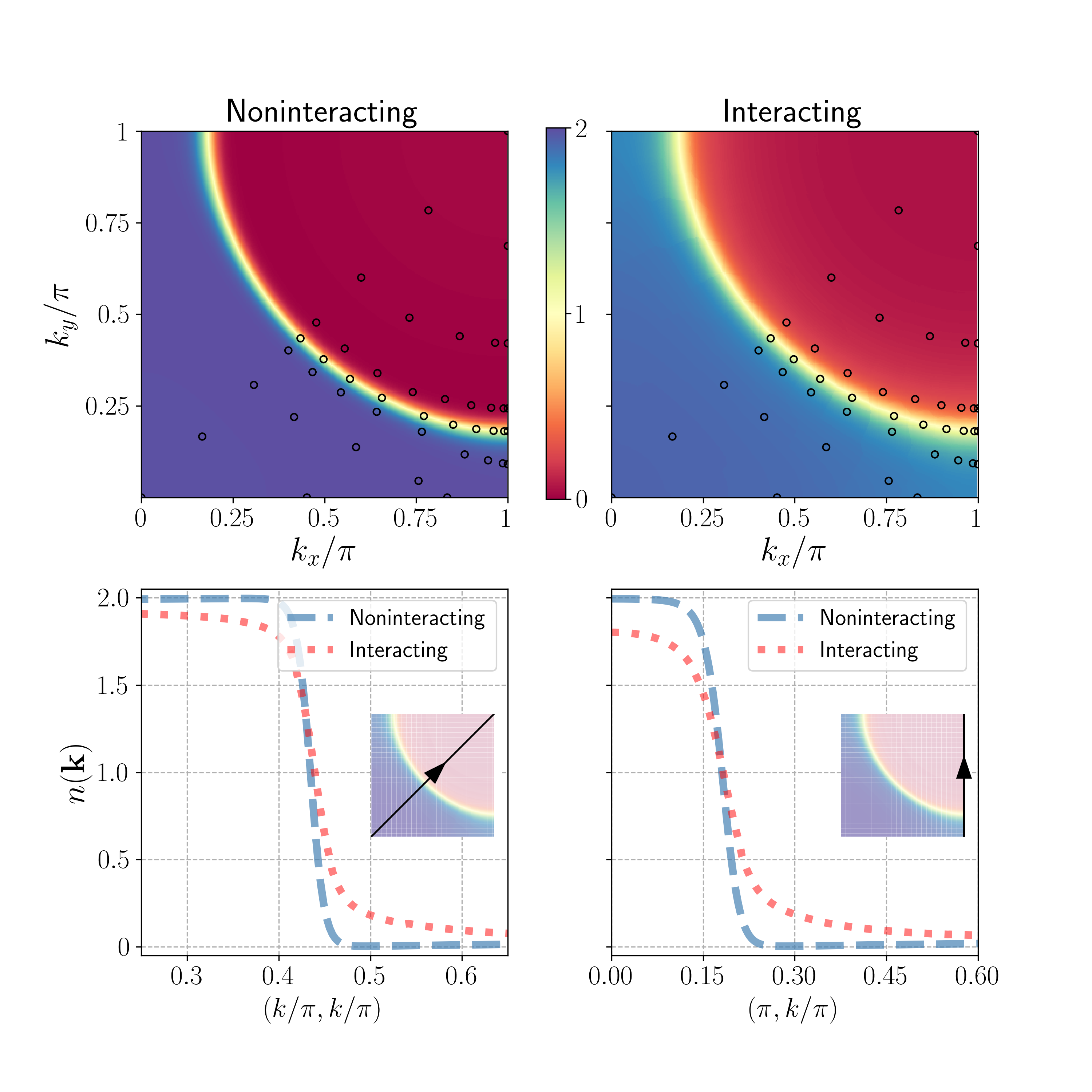}
\caption{Top row: momentum distribution for $t'=-0.32t$, $T=0.08t$ and doping $x=0.025$. Left panel: non-interacting case. Right panel: interacting case for $U=4t$. The black circles mark the points used to patch the self-energy.
Bottom row: cut of the occupation along the Brillouin zone paths reported as arrows in the insets. Blue dashed curves are results for the non-interacting system, while red dotted curves are for $U=4t$. } 
\label{fig:occ975}
\end{figure}

\begin{figure}
\includegraphics[width=0.5\textwidth]{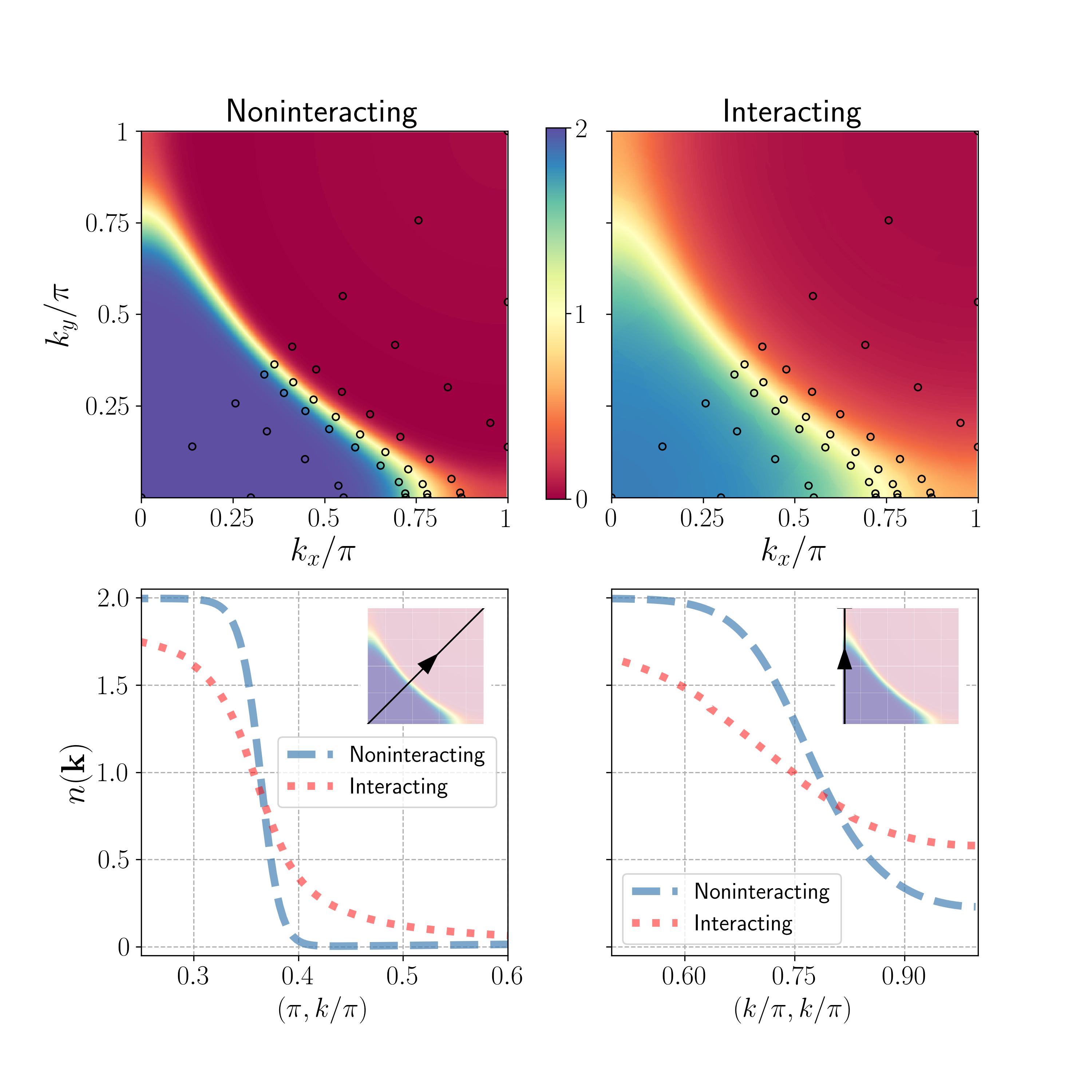}
\caption{Top row: momentum distribution for $t'=-0.32t$, $T=0.08t$ and doping $x=0.4$. Left panel: non-interacting case. Right panel: interacting case for $U=4t$. The black circles mark the points used to patch the self-energy.
Bottom row: cut of the occupation along the Brillouin zone paths reported as arrows in the insets. Blue dashed curves are results for the non-interacting system, while red dotted curves are for $U=4t$. }
\label{fig:occ600}
\end{figure}

The suppression of instabilities, and in particular the elimination of the artificial charge instability by dynamical self-energy feedback was already observed by Husemann et al.\ \cite{Husemann2012} In that work, however, the momentum dependence of the self-energy was approximated by its value at the van Hove points, where it is particularly large. The suppression effects are thereby likely somewhat overestimated.

Trying to understand the self-energy feedback effects, we looked for possible changes in the Fermi surface shape by analyzing the momentum distribution~\cite{Note4}

\begin{equation}
 n^{\Lambda}(\mathbf{k})  = 2T \sum_{\nu}\frac{e^{i\nu 0^+}}{i\nu-\varepsilon_\bs{k}+\mu^\Lambda-\Lambda\Sigma^\Lambda(\bs{k},\nu)}.
 \label{eq:occ} 
\end{equation}

In Fig.~\ref{fig:occ975} we show the non-interacting (top left) and interacting (top right) momentum distribution in the first quadrant of the Brillouin zone for doping $x=0.025$.
The latter is computed at the critical scale $\Lambda_\mathrm{c}$. 
Comparing the two panels, one does not observe any relevant shift of the Fermi surface position, but the Fermi surface broadening is appreciably larger in the interacting case, due to the self-energy.
Similar results apply for doping $x=0.4$, as one can see from Fig.~\ref{fig:occ600}, where the broadening is more evident.  

\begin{figure}
\includegraphics[width=0.5 \textwidth]{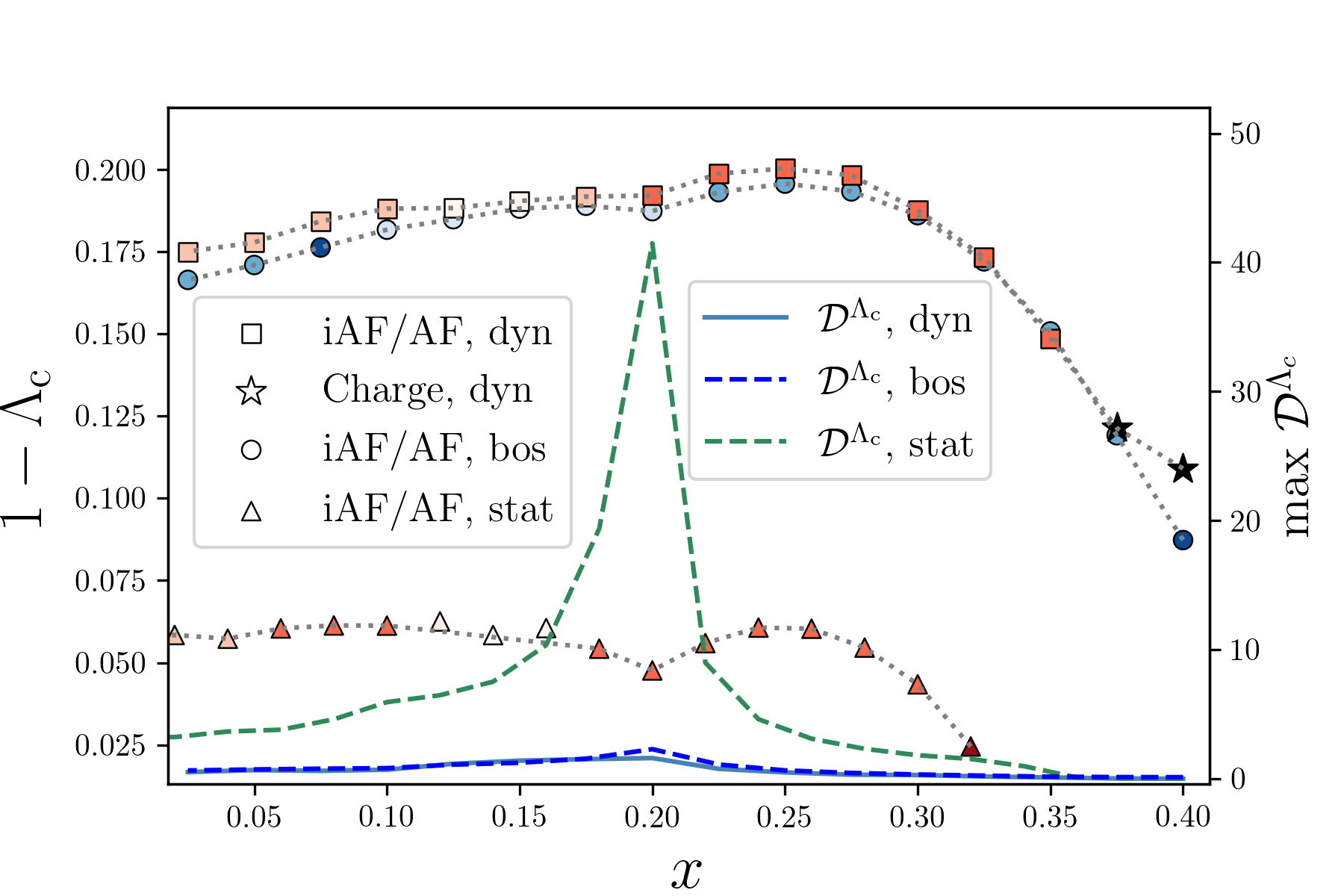}
\caption{Critical scale as a function of doping $x=1-n$, for $T = 0.08t$, $t'=-0.32t$ and $U=4t$. 
Squares, circles and triangles refer to leading couplings in the magnetic channel for dynamic, bosonic, and static implementations respectively. The black stars refer to a divergence in the charge channel at $\mathbf{Q} = (0,0)$. In all the implementations, no self-energy feedback has been used.
The color of squares and circles encodes the distance of the incommensurate magnetic $\mathbf{Q}$-vector from $(\pi,\pi)$: darker color corresponds to a larger distance, as in Fig.~\ref{fig:criscale}.
The maximal value of $\mathcal{D}^\Lambda$ at the critical scale is marked by a solid blue line for the dynamic implementation, by a dashed light blue line for the bosonic approximation, and by a dashed green line for the static approximation.
}  
\label{fig:PDstatic}
\end{figure} 

\begin{figure}
\hspace*{-1.0cm}
\includegraphics[width=0.48\textwidth]{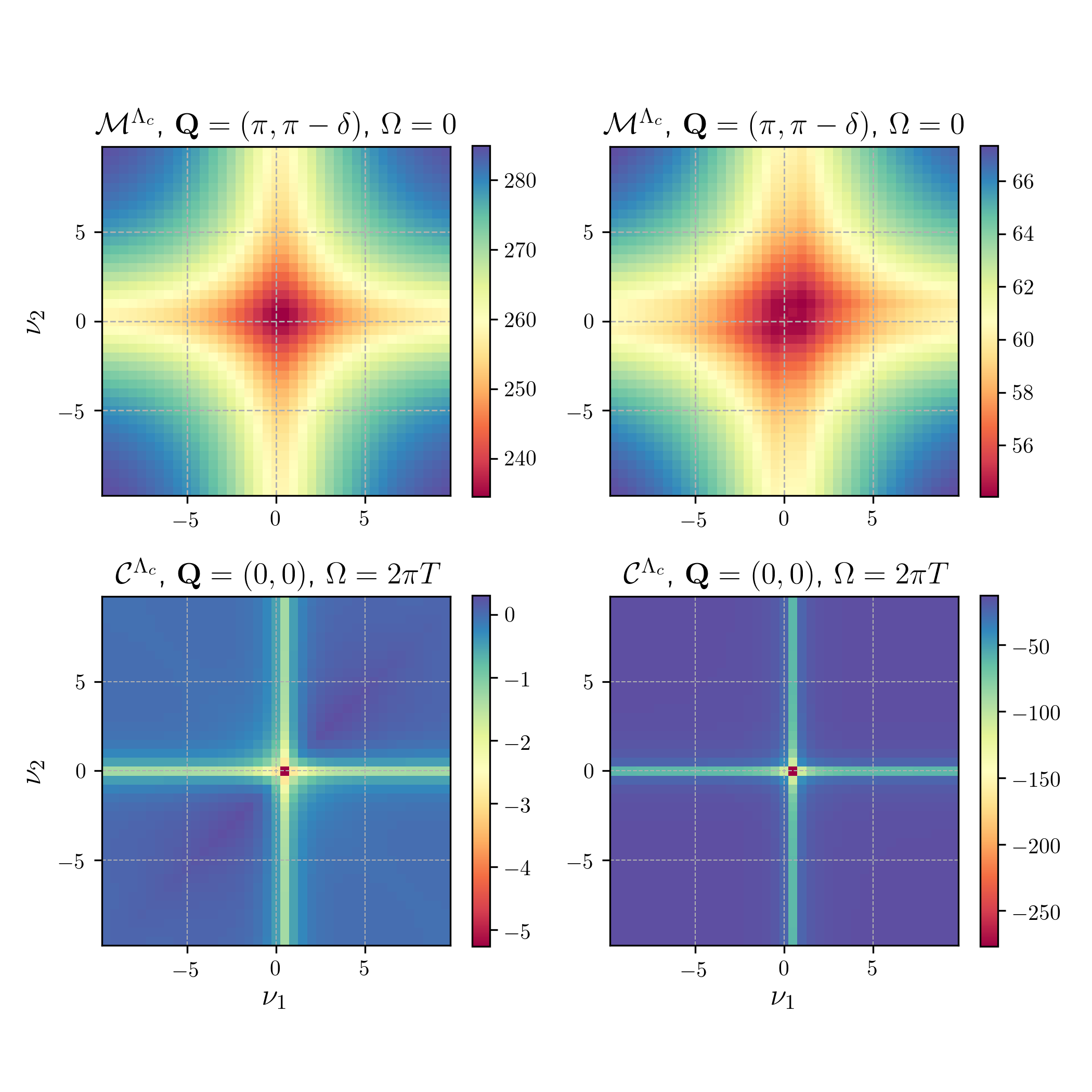}
\caption{Frequency dependence of the magnetic (top) and charge (bottom) channel for $t'=-0.32$, $U=4t$ and $T=0.08t$.
\emph{Top left}:
Magnetic channel $\mathcal{M}^\Lambda_{\bs{Q},\Omega}(\nu_1,\nu_2)$ with self-energy feedback at the instability vector and for vanishing frequency transfer, for doping $x=0.025$.
\emph{Top right}:
Magnetic channel $\mathcal{M}^\Lambda_{\bs{Q},\Omega}(\nu_1,\nu_2)$ without self-energy feedback at the instability vector and for vanishing frequency transfer, for doping $x=0.4$.
\emph{Bottom left}:
Frequency dependence of the charge channel $\mathcal{C}^\Lambda_{\bs{Q},\Omega}(\nu_1,\nu_2)$ with self-energy feedback at $\mathbf{Q}=(0,0)$ and frequency transfer $\Omega=2\pi T$, for doping $x=0.025$.
\emph{Bottom right}:
Frequency dependence of the charge channel $\mathcal{C}^\Lambda_{\bs{Q},\Omega}(\nu_1,\nu_2)$ without self-energy feedback at $\mathbf{Q}=(0,0)$ and frequency transfer $\Omega=2\pi T$, for doping $x=0.4$.
}  
\label{fig:freqplot} 
\end{figure}

In Fig.~\ref{fig:PDstatic}, we compare the critical scales for different approximations on the vertex frequency dependence: our fully dynamic approach, a \emph{bosonic} approximation scheme (with a separable frequency dependence of the vertex), and a static approximation.
In none of these results we took the self-energy feedback into account.
The static approximation is obtained by completely neglecting the frequency structures of the channels, assuming the vertex to be constant in frequency space. 
Following Ref.~\onlinecite{Husemann2009}, we evaluate the flow equations only for $\Omega=0$, as transfer frequency, and $\pm \pi T$ as fermionic arguments. 
Instead, in the bosonic scheme we let each channel depend on the transfer frequency only, as motivated by perturbative arguments indicating a weak dependence on the other two frequencies.\cite{Karrasch2008,Husemann2012}
However, as shown in the next section, already at moderate coupling the effective interactions have strong dependences on the other two frequency arguments, too.
For this reason, there is an ambiguity in the way the interaction channels are projected to a function of a single bosonic frequency.  
Different projection schemes lead to quantitatively different results.
In Fig.~\ref{fig:PDstatic} we show the results from a low-frequency projection that leads to the critical scale most consistent with the one of the fully frequency dependent scheme.

We observe that $\Lambda_{\mathrm{c}}$ is higher in the static case, that is, the instability occurs at a larger $U^\Lambda$. This is due to two reasons, first, by taking $\nu_1=-\nu_2=\pi T$ the leading magnetic channel (at fixed bosonic frequency) is approximated by its minimal value, as will be shown in the next paragraph. Second, in the static approximation the feedback of the other channels is overestimated, see below.
For $x \ge 0.34$ there is no divergence in any channel for the temperature considered. 
  
In Fig.~\ref{fig:PDstatic} we also show the maximal value of the $d$-wave pairing interaction $\mathcal{D}^\Lambda$ at $\Lambda_c$ in the static, bosonic and fully dynamic parametrizations.
In none of these results $d$-wave pairing is the leading instability at the temperature under consideration, but in the static approximation $\mathcal{D}^{\Lambda_{\mathrm{c}}}$ is orders of magnitude larger than in the other two cases.
At lower temperatures (not shown here) we do observe a $d$-wave pairing instability in the static approximation.
This suppression of pairing by the frequency dependence of the vertex, already observed by Husemann et al.,\cite{Husemann2012} has been addressed above in this section.


\subsection{Frequency dependence of vertex}

We now discuss the remarkable frequency dependence of the vertex. 
We will first look at the channels showing a divergence, that is, the charge and the magnetic instabilities observed in Fig.~\ref{fig:criscale}, and we will then discuss the pairing channels.

As mentioned in the previous section, the divergences of the charge and magnetic channels are quite different. The charge channel diverges for a finite frequency transfer, and only when we neglect the self-energy feedback. 
Since the dependence on the transfer momentum and frequency $(\bs{Q},\Omega$) has already been discussed in Ref.~\onlinecite{Husemann2012}, we focus on the dependence on the fermionic frequencies. Therefore we present various color plots for fixed  $(\mathbf{Q},\Omega)$, showing the dependence on $\nu_1$ and $\nu_2$. 
 
In the top left panel of Fig.~\ref{fig:freqplot} we show the magnetic channel $\mathcal{M}^{\Lambda_\mathrm{c}}_{\mathbf{Q},\Omega}(\nu_1,\nu_2)$ in the small doping region, where antiferromagnetism is the leading instability. 
The results shown have been calculated with self-energy feedback, but the frequency structures we discuss do not depend strongly on the presence of the self-energy. 
For clarity we restrict the plots to the first $20$ positive and negative Matsubara frequencies.
When only one channel in Eq.~(\ref{eq:vertflow}) is taken into account, the fRG equations are equivalent to the RPA . 
The magnetic channel calculated with RPA would depend only on the frequency and momentum transfer.
Hence any variation in the frequency structure has to be ascribed to the presence of the other channels in the fRG.
The channel competition suppresses the magnetic channel: the largest value of $\mathcal{M}$ is reduced compared to the RPA, and the frequency dependent structure at the center is further reduced compared to the asymptotic values at large $\nu_1$, $\nu_2$. 

In the bottom left panel of Fig.~\ref{fig:freqplot} we show the frequency dependence of the charge channel $\mathcal{C}^{\Lambda_\mathrm{c}}_{\bs{Q},\Omega}(\nu_1,\nu_2)$ for a finite frequency transfer $\Omega=2\pi T$, related to the charge instability discussed in Ref.~\onlinecite{Husemann2012} and above. 
The frequency structure is completely different from the magnetic channel. 
The charge channel has its maximum for frequencies $\nu_1 = \pi T$ and $\nu_2=-\pi T$. 
This structure cannot be explained in terms of standard ladder diagrams. It might be related to the behavior of the retarded interaction described in Ref.~\onlinecite{Hafermann2014,*Stepanov2016}.
In the two right panels of Fig.~\ref{fig:freqplot} we show the same quantities but for $x=0.4$ and without self-energy feedback. In this case, the localized peak in the charge channel is the leading interaction. The position and shape of the frequency structures are similar to the one described above.

\begin{figure}[tbh]
    \includegraphics[scale=0.55]{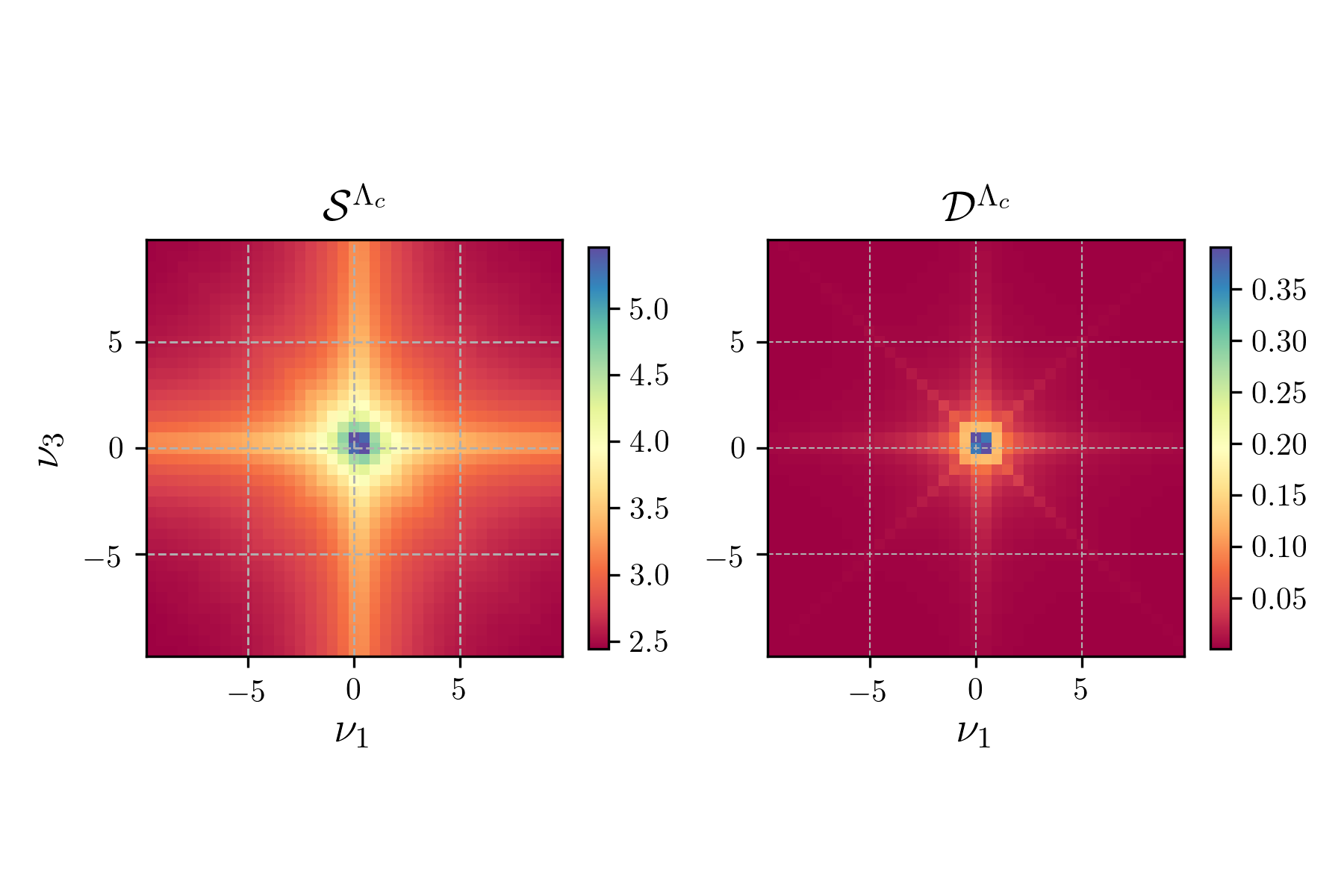} \\[-0.8cm]
    \includegraphics[scale=0.55]{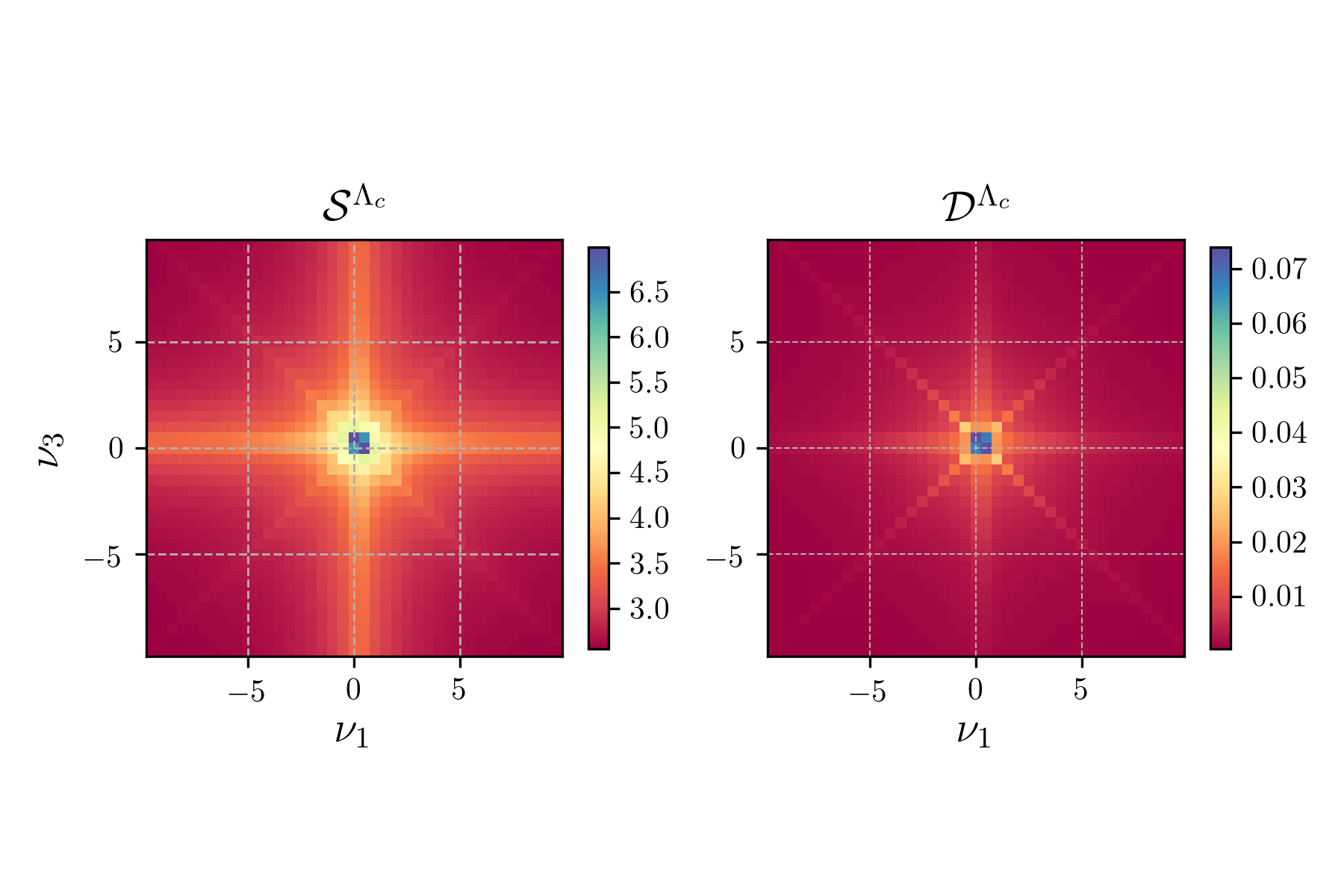}
\caption{Frequency dependence of the pairing channels
 $\mathcal{S}^{\Lambda_c}_{\bs{Q},\Omega}(\nu_1,\nu_3)$ and
 $\mathcal{D}^{\Lambda_c}_{\bs{Q},\Omega}(\nu_1,\nu_3)$ for $\bs{Q}=(0,0)$ and $\Omega=0$.
 The doping is $x=0.025$ (top) and $x=0.4$ (bottom). The other parameters are $T=0.08t$, $t'=-0.32t$, and $U=4t$.}
\label{fig:pairing}
\end{figure}

In Fig.~\ref{fig:pairing} we display the frequency dependence of the pairing functions $\mathcal{S}$ and $\mathcal{D}$ for two distinct doping values $x=0.025$ and $x=0.4$.
One can see that $\mathcal{D}^{\Lambda_c}$ is indeed asymptotically vanishing at large frequencies,\cite{Wentzell2016a} as can be understood from the frequency dependences in Eqs.~(\ref{eq:dwaveflow}) and (\ref{eq:Ldwave}).

\begin{figure*}
\includegraphics[width=\textwidth]{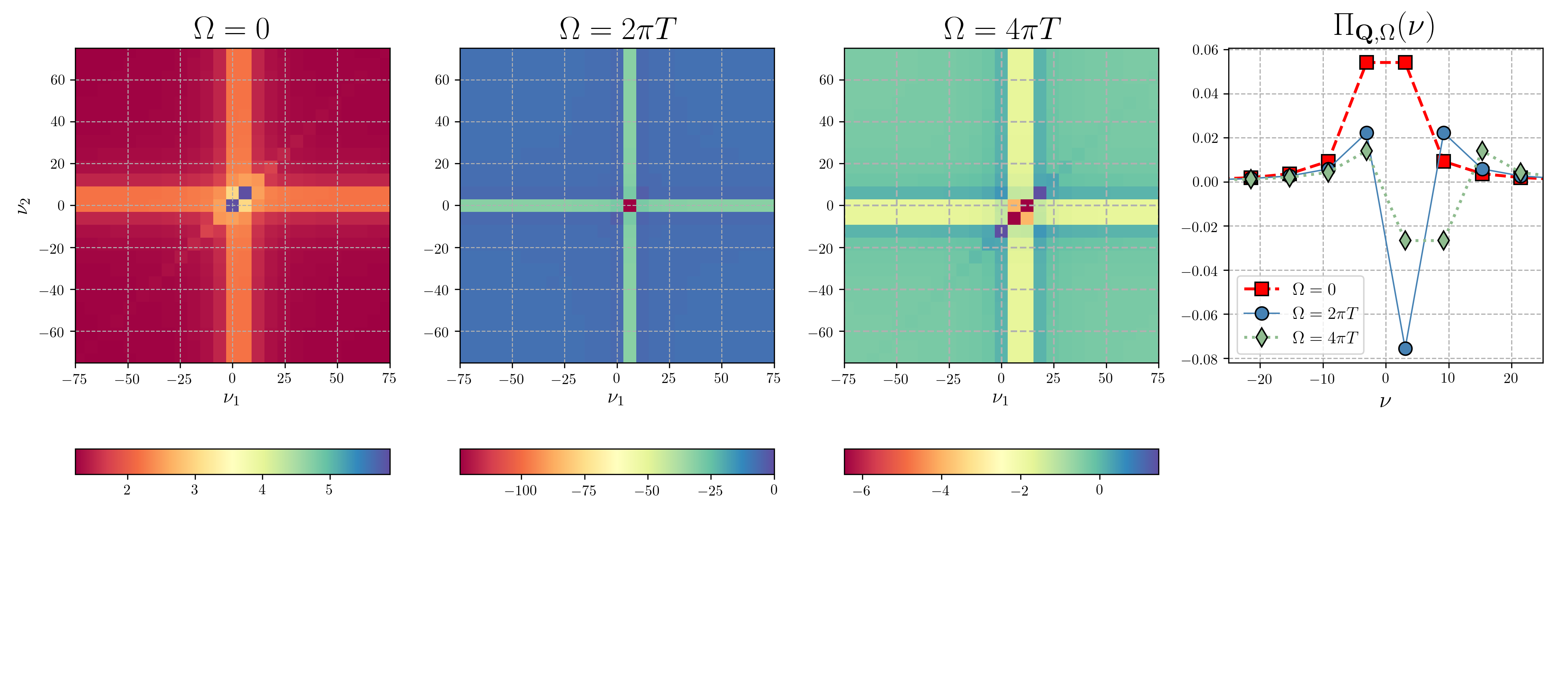}
\vspace*{-2.0cm}
\caption{In the first three panels from the left, the charge channel
$\mathcal{C}_{\bs{Q},\Omega} (\nu_1,\nu_2) =
 \mathcal{\tilde C}_{\bs{Q}=(0,0),\Omega}(\nu_1,\nu_2-\Omega)$
computed from Eq.~(\ref{pl:charge}) is shown as a function of $\nu_1$ and $\nu_2$ for transfer frequencies $\Omega=0$, $\Omega=2\pi T$ and $\Omega=4\pi T$, respectively. In the right panel, the bubble $\Pi_{\bs{Q}=(0,0),\Omega}(\nu)$ is shown as a function of $\nu$ for $\Omega=0$, $\Omega=2\pi T$ and $\Omega=4\pi T$. The model parameters are $t'=-0.32$ and $U=4$, the doping $x=0.375$, and the temperature $T=t$.} 
\label{fig:perpladder}
\end{figure*}


\subsection{Origin of charge singularity} 
\label{sec:PerpLadder}

To gain insight into the origin of the singular frequency structures observed in the charge channel, we identify a simple set of Feynman diagrams reproducing the same features.   
The main idea is that the magnetic channel, which is generated first, is responsible for the singular structure in the charge channel. 

To check this qualitatively, we first compute an effective interaction by means of an RPA in the magnetic channel, and then insert this effective magnetic interaction into a subsequent RPA equation for the charge channel. 
Of course one does not expect quantitative agreement with the fRG, since we overestimate both interactions, but the approximation is sufficient to reproduce and explain the qualitative features we are interested in.  
  
We start by introducing an effective interaction that includes the magnetic fluctuations as computed by RPA in the particle-hole crossed channel:
\begin{equation}
 U^{\mathrm{eff}}_{\bs{Q},\Omega} = \frac{U}{1 - U \Pi_{\bs{Q},\Omega}}.
\label{pl:ueff}
\end{equation}
Since the bare interaction $U$ is local, $U^{\mathrm{eff}}$ depends only on the transfer momentum $\bs{Q}$ and frequency $\Omega$ of the particle-hole bubble
\begin{equation}
 \Pi_{\bs{Q},\Omega} =
 - T \sum_{\nu} \int_{\bs{p}} G_0(\bs{p},\nu) G_0(\bs{p}+\bs{Q},\nu+\Omega).
\end{equation}
The magnetic effective interaction in Eq.~(\ref{pl:ueff}) will now be used to compute the RPA equation for the charge channel. Adopting the simplified momentum dependences of the effective interactions used in the fRG calculation, only the momentum integrated, that is, local part of the magnetic interaction
$U^{\mathrm{eff}}_{\Omega} = \int_{\bs{Q}} U^{\mathrm{eff}}_{\bs{Q},\Omega}$
contributes to the charge channel. We thus obtain
$\mathcal{C}_{\bs{Q},\Omega} (\nu_1,\nu_2) =
 \mathcal{\tilde C}_{\bs{Q},\Omega} (\nu_1,\nu_2 - \Omega)$, where
\begin{equation}
 \mathcal{\tilde C}_{\bs{Q},\Omega} (\nu_1,\nu_3) = -
 U^{\mathrm{eff}}_{\nu_1-\nu_3} \left[ \delta_{\nu_1,\nu_3} + U^{\mathrm{eff}}_{\nu_1-\nu_3}
 \Pi_{\bs{Q},\Omega}(\nu_1) \right]^{-1} ,
 \label{pl:charge}
\end{equation}
with
\begin{equation}
 \Pi_{\bs{Q},\Omega}(\nu) =
 - T \int_{\bs{p}} G_0(\bs{p},\nu) G_0(\bs{p}+\bs{Q},\nu+\Omega) .
\end{equation}
Note that the fermion frequencies $\nu$ are not summed in $\Pi_{\bs{Q},\Omega}(\nu)$, and the inverse in Eq.~(\ref{pl:charge}) is a matrix inverse of the matrix with indices $\nu_1$ and $\nu_3$.
Eq.~(\ref{pl:charge}) is nothing but an RPA equation with a frequency dependent interaction in the particle-hole channel. $U^{\mathrm{eff}}$ depends on $\nu_1-\nu_3$ due to the frequency exchange from particle-hole crossed to particle-hole notation.
In the case of a frequency independent effective interaction $U^{\mathrm{eff}}$, Eq.~(\ref{pl:charge}) becomes $\nu_1$ and $\nu_3$ independent 
and only the summed bubble $\Pi_{\bs{Q},\Omega}$ appears.
The frequency dependence of $U^{\mathrm{eff}}$ qualitatively affects the results. 

In Fig.~\ref{fig:perpladder}, we show the charge channel as computed from Eq.~(\ref{pl:charge}) for $\bs{Q}=(0,0)$ and different $\Omega$ as a function of $\nu_1$ and $\nu_2 = \nu_3 + \Omega$, for $T=t$ and $x=0.375$.
We have to choose such a high temperature to stay in a stable paramagnetic phase, due to the above-mentioned overestimation of the fluctuations within the RPA. In the more accurate fRG calculation the magnetic instability occurs at lower temperatures.
The frequency structure in Fig.~\ref{fig:perpladder} for $\Omega=2\pi T$ is very similar to the one shown in Fig.~\ref{fig:freqplot}. 
The simple contributions considered here reproduce the position of the main structures, as well as the correct sign of the charge channel. 
This is true also for the other bosonic Matsubara frequencies shown here, for which we do not report the fRG results. 
Furthermore, upon lowering the temperature the charge channel diverges also for other finite bosonic Matsubara frequencies, while it does not diverge for $\Omega=0$.
From this we conclude that the frequency dependent effective magnetic interaction described above is responsible for the frequency structure of the charge channel observed in the fRG. 

To understand why the divergence appears for a non-zero frequency $\Omega$, we notice that in Eq.~(\ref{pl:charge}) the $\Omega$ dependence 
appears only through the bubble $\Pi_{\bs{Q},\Omega}(\nu)$. The frequency summed particle-hole bubble obeys the following relation:  
\begin{equation}
 \Pi_{\bs{Q}\rightarrow(0,0),\Omega} = \sum_{\nu}
 \Pi_{\bs{Q}\rightarrow(0,0),\Omega}(\nu) = C \delta_{\Omega,0},
\label{pl:sumrule}
\end{equation}
where $C$ is a positive constant that, at low temperature, approaches the density of states at the Fermi level.
In the rightmost panel of Fig.~\ref{fig:perpladder}, we show the bubble $\Pi_{\bs{Q}=(0,0),\Omega}(\nu)$ as a function of $\nu$ for different values of $\Omega$.
We note that it has a large negative peak for $\Omega=2\pi T$.
This is due to the property (\ref{pl:sumrule}): the summed bubble must vanish for $\Omega \neq 0$, hence a large negative value is needed to cancel the positive contributions at large frequency. 
We have thus identified the origin of the frequency structure observed in the charge channel, which seems to be quite general since arises from simple Feynman diagrams. 

Including the self-energy in the calculation of the bubble,   Eq.~(\ref{pl:sumrule}) does not evaluate to a $\delta$-function anymore, and the difference between the summed bubble at vanishing frequency and for frequency $2\pi T$ is diminished. 
This is probably the reason why the inclusion of the self-energy feedback prevents the unphysical divergence of the charge channel.     

\subsection{Self energy}

\begin{figure*}[tbh]
    \subfigure[$x=0.025$]{
    \label{fig:selffermi0975}
    \includegraphics[scale=0.55]{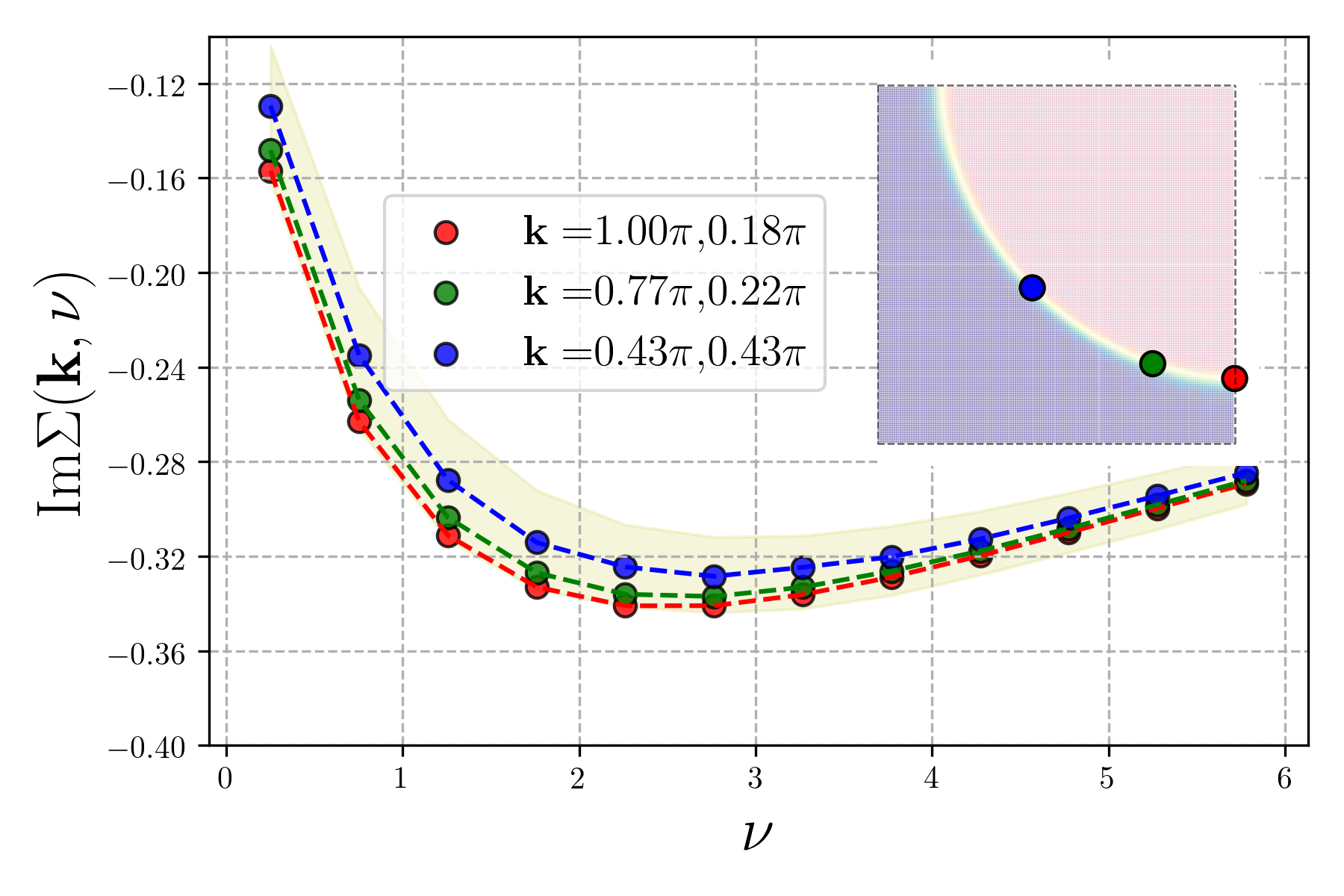}}
    \subfigure[$x=0.400$]
    {
    \label{fig:selffermi0600}
    \includegraphics[scale=0.55]{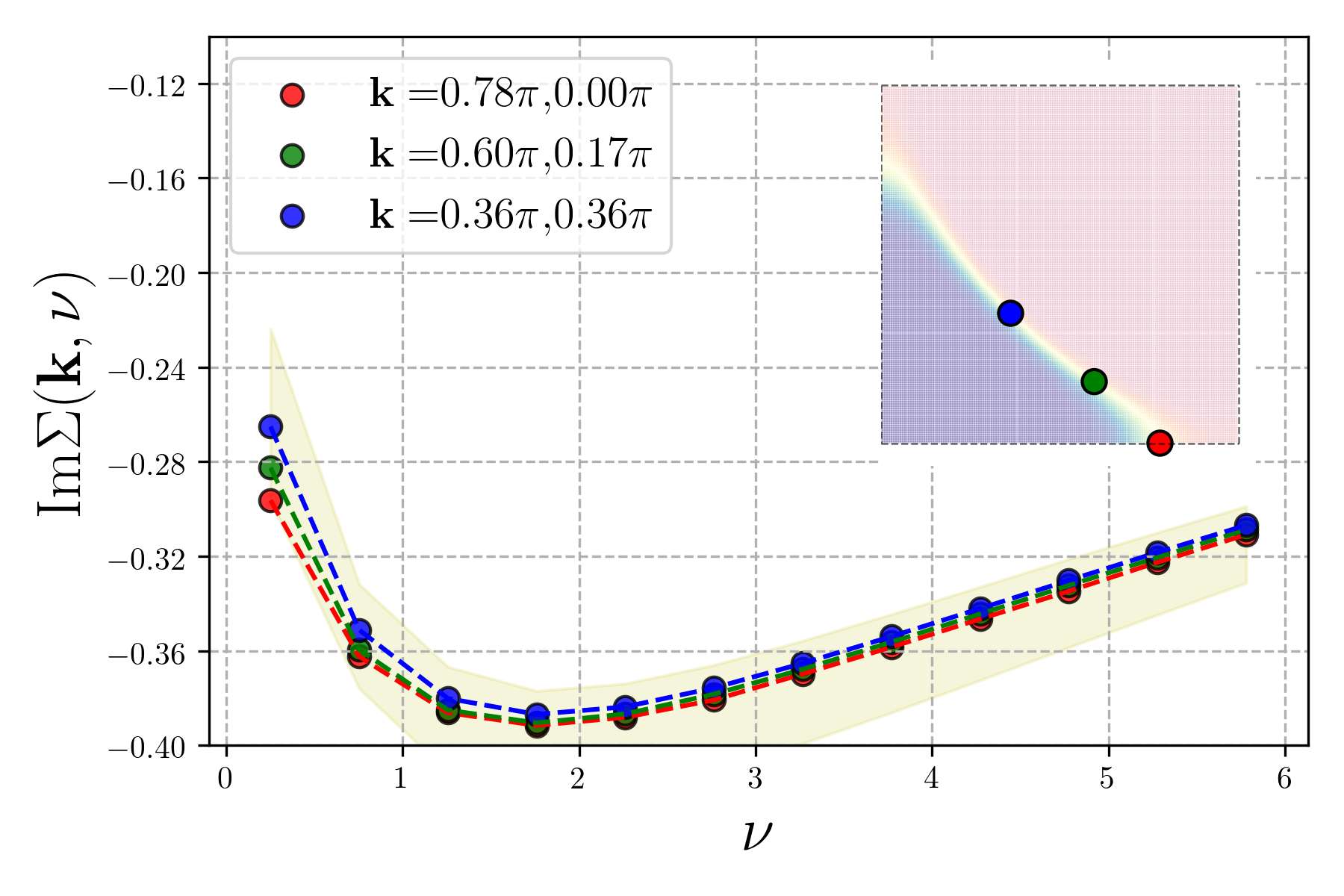}} 
  \caption{Self-energy as a function of frequency for $U=4t$, $t'=-0.32t$ at  temperature $T=0.08t$.
The location of the $\mathbf{k}$-point in the Brillouin zone is color coded in the inset. The position of all the patching points taken into account for the self-energy is shown as black circles in the top row of Figs.~\ref{fig:occ975} and \ref{fig:occ600}, and does not change during the flow.
The shaded area highlights the region between the maximal and minimal value of the self-energy for each frequency. }
  \label{fig:selffermi}
\end{figure*}
\begin{figure}
\includegraphics[width=0.5\textwidth]{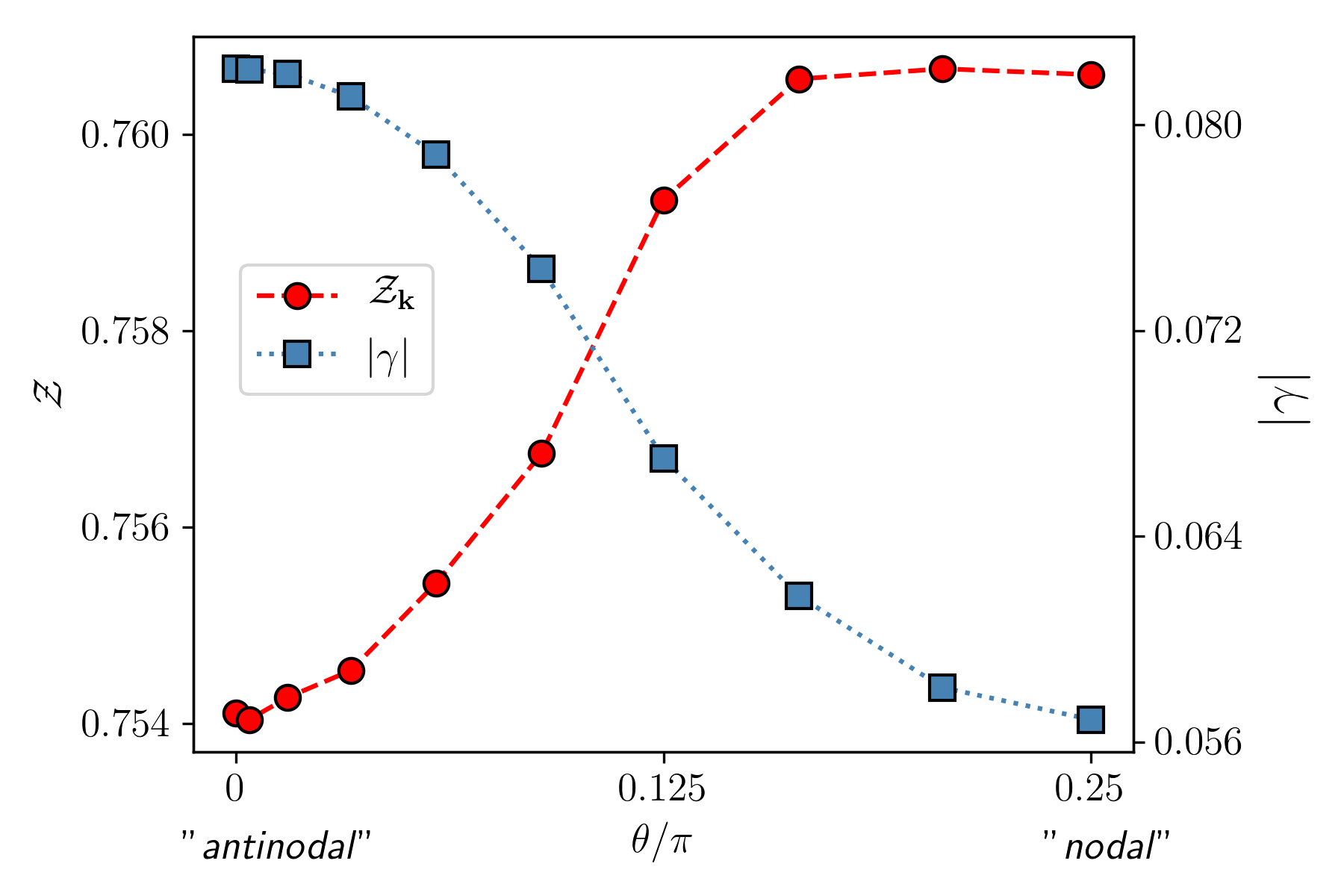}
\caption{Quasiparticle weight $\mathcal{Z}_{\bs{k} }$ and decay rate $\gamma_{\bf{k}}$ as function of the angle $\theta$ for the same parameters as in Fig.~\ref{fig:selffermi0975}. 
The values on the left axis refer to the quasiparticle weight, the values on the right axis refer to the decay rate.}
\label{fig:zetaandgamma}
\end{figure}

We now discuss the frequency and momentum dependence of the self energy. 
In Fig.~\ref{fig:selffermi0975}  we show the frequency dependence of the imaginary part of the self-energy at $T=0.08t$ and low doping $x=0.025$. 
The spread between the maximal and minimal self-energy at each frequency is rather small, indicating that the self-energy did not develop a large momentum dependence even when the flow parameter reached the critical scale. 
At small frequencies the self-energy has a typical Fermi liquid behavior.
One would generally expect the antinodal region to be more affected by correlation effects. However, there is only a slight increase of $|\mathrm{Im}\Sigma(\bs{k},\nu)|$ in this region. At the temperature 
and interaction strength we are considering, we do not observe a tendency towards the opening of a momentum selective gap. 
In Fig.~\ref{fig:selffermi0600} we show the imaginary part of the self-energy for a larger doping $x=0.4$. As in the previous case, we do not see much momentum differentiation.

The self-energy enters directly in the calculation of the momentum distribution through the Green's function, already discussed above, and shown in Figs.~\ref{fig:occ975} and~\ref{fig:occ600}.
In the bottom panels of these figures, we show how the momentum distribution evolves along two different cuts in the Brillouin zone, crossing the \textit{nodal} and \textit{antinodal} regions, respectively.
The drop in the momentum distribution is sharper along the diagonal, and the self-energy effects are stronger along the antinodal cut.
For doping $x=0.4$ the broadening of the Fermi surface, already larger at the non interacting level, is further enhanced by the self-energy.

To study further the difference between nodal and antinodal regions in the iAF regime, we studied the quasiparticle weight \cite{Abrikosov1963} $\mathcal{Z}_{\bs{k}}$, and the decay rate $\gamma_{\bs{k}}$.
Instead of relying on analytical continuation, we have extracted the parameters directly from the Matsubara frequencies data.
To do so we have fitted the first few frequencies of the imaginary part of the self-energy with a polynomial of degree $l$: $\mathrm{Im}\Sigma(\bs{k},\nu) \approx a_0(\bs{k})+a_1(\bs{k})\nu+...+a_l(\bs{k})\nu^l$ and we identified $\gamma_{\mathbf{k}}=a_0(\bs{k})$ and $\mathcal{Z}_{\bs{k}}= [1-a_1(\bs{k})]^{-1}$.
The procedure only works if the temperature is low enough, and if the frequencies used for the fit are not too high. We checked that the results were stable upon changing the number of frequencies and the order of the polynomial used for the fit. 
In Fig.~\ref{fig:zetaandgamma} we plot $\mathcal{Z}_{\bs{k}}$ and $\gamma_{\bs{k}}$ against the angle $\theta$ along the Fermi surface, $\theta=0$ corresponding to the antinodal direction and $\theta=\pi/4$ to the nodal one. 
The variation of the quasiparticle weight along the Fermi surface is extremely small with $\mathcal{Z}_{\bs{k}}$ assuming values between $0.754$ and $0.760$. 
On the other hand, the relative variation of the decay rate $\gamma$ along the Fermi surface is sizable, varying from $\gamma\approx 0.056t$ to $\gamma \approx 0.082t$. These values are comparable with the temperature $T=0.08t$.

Decay rates \cite{Honerkamp2001a} and quasi-particle weights \cite{Honerkamp2003} were computed already in early fRG calculations from two-loop contributions to the self-energy, obtained by inserting the integrated one-loop equation for the vertex into the flow equation for the self-energy. In this way the computation of a frequency dependent vertex was avoided.
The size and anisotropy of the decay rates obtained in these calculations are comparable to our results. The quasi-particle weight was even less reduced, and its anisotropy more pronounced, probably because the Fermi surface in Ref.~\onlinecite{Honerkamp2003} is more nested than ours and close to van Hove points. 

We conclude that near the critical scale the system generically still has coherent quasiparticles along the Fermi surface, with a higher decay rate in the antinodal region. This is consistent with the results of Ref.~\onlinecite{Katanin2004,*Rohe2005}, where non-Fermi liquid behavior of the self-energy was observed only very close to the pseudo-critical temperature and in the immediate vicinity of the magnetic hot spots.


\section{Conclusions}
\label{sec:conclusions}

We have applied fRG flow equations to the two-dimensional Hubbard model, using a form factor decomposition for the momentum arguments of the two-particle vertex, but maintaining intact all the frequency dependencies with a high resolution. 

The frequency dependence tends to enhance magnetic fluctuations and suppress $d$-wave pairing fluctuations. These tendencies are in agreement with previous results obtained from
an approximate separable ansatz for the frequency dependence of the vertex. \cite{Husemann2012}
The complexity of the fully frequency dependent implementation is rewarded by the possibility of accessing and understanding the frequency structures arising in the flow.
We confirm that, in a flow without self-energy feedback, there exist regions of parameter space where the vertex shows a divergence in the charge channel at non-zero frequency, as already found by Husemann et al.\cite{Husemann2012}
We are able to identify a simple set of Feynman diagrams that give rise to the above-mentioned divergence, which are likely to generate unexpected singular features in the charge channel also in other theories that take into account both the frequency dependence of the vertex and the interplay of different fluctuation channels.\cite{Stepanov2016}
 
The proper treatment of the frequency dependence of the vertex allows for a calculation of the frequency dependent self-energy.
We observed that the feedback of the self-energy into the vertex flow plays an important role, also at the qualitative level, since it suppresses the unphysical divergence in the charge channel. 

Given the increasing importance of the frequency dependence as more correlated regimes are approached, our work paves the way for future developments of the fRG for correlated fermion systems.
At moderate coupling, like the one treated here, the combination of a frequency dependent vertex and self-energy feedback allows to revisit and improve previous results. At strong coupling, a non-perturbative starting point is needed. 
This is what is proposed in DMF$^2$RG,\cite{Taranto2014} where the flow starts from the DMFT solution for the vertex and the self-energy, which are both strongly frequency dependent. Therefore, consistently taking into account the frequency dependence is crucial to access strongly interacting fermion systems.


\vspace*{5mm}
\begin{acknowledgments} 
We are grateful to M.~Salmhofer, A.~Eberlein, S.~Andergassen, C.~Honerkamp, and A.~Toschi for useful discussions. 
We thank O.~Gunnarsson for a critical reading of the manuscript and D.~T.~Mantadakis for comments and suggestions. 
\end{acknowledgments}


\begin{appendix}

\begin{widetext}

\section{Flow equations}
\label{sec:FlowEquations}

Here we present the final expressions for the flow equations in the pairing and in the charge channels. The flow equations for the magnetic channel have been presented in
Sec.~\ref{sec:vertex}.

The flow equation for the $s$-wave pairing channel reads
\begin{equation}
\dot{\mathcal{S}}_{\bs{Q},\Omega}(\nu_1,\nu_3) = 
  T \sum_\nu L^{\mathrm{s},\Lambda}_{\mathbf{Q},\Omega}(\nu_1,\nu) P^{\mathrm{s},\Lambda}_{\bs{Q},\Omega}(\nu)
  L^{\mathrm{s},\Lambda}_{\mathbf{Q},\Omega}(\nu,\nu_3),
\end{equation}  
with
\begin{equation}
 P^{\mathrm{s},\Lambda}_{\bs{Q},\Omega}(\omega) = \int_{\bs{p}}
 G^{\Lambda}(\bs{p},\omega)S^{\Lambda}(\bs{Q}-\bs{p},\Omega-\omega) +
 G^{\Lambda}(\bs{Q}-\bs{p},\Omega-\omega) S^{\Lambda}(\bs{p},\omega), 
\label{eq:app:P_pp_s}
\end{equation} 
and
\begin{align} 
\label{eq:Lswave}
 L^{\mathrm{s},\Lambda}_{\mathbf{Q},\Omega}(\nu_1,\nu_3) = U-\mathcal{S}^{\Lambda}_{\bs{Q},\Omega} (\nu_1,\nu_3) +
 \int_{\bs{p}} \Big[ \mathcal{M}^{\Lambda}_{\bs{p},\nu_3-\nu_1}(\nu_1,\Omega-\nu_1) + \frac{1}{2} \mathcal{M}^{\Lambda}_{\bs{p},\Omega-\nu_1-\nu_3}(\nu_1,\Omega-\nu_1) -
 \frac{1}{2} \mathcal{C}^{\Lambda}_{\bs{p},\Omega-\nu_1-\nu_3}(\nu_1,\Omega-\nu_1) \Big]. 
\end{align}	 
The flow equation for the $d$-wave pairing channel reads
\begin{equation}
 \dot{\mathcal{D}}^{\Lambda}_{\bs{Q},\Omega}(\nu_1,\nu_3) = 
 T \sum_\nu L^{\mathrm{d},\Lambda}_{\bs{Q},\Omega}(\nu_1,\nu)
 P^{\mathrm{d},\Lambda}_{\bs{Q},\Omega(\nu)}
 L^{\mathrm{d},\Lambda}_{\bs{Q},\Omega} (\nu,\nu_3), 
\label{eq:dwaveflow}
\end{equation}
with
\begin{equation}
 P^{\mathrm{d},\Lambda}_{\bs{Q},\Omega}(\omega) =
 \int_{\bs{p}} \left[ f_{\mathrm{d}}\left( \bs{Q}/2 - \bs{p} \right) \right]^2 
\left[ G^{\Lambda}(\bs{p},\omega)S^{\Lambda}(\bs{Q}-\bs{p},\Omega-\omega) +G^{\Lambda}(\bs{Q}-\bs{p},\Omega-\omega)
S^{\Lambda}(\bs{p},\omega) \right], 
\label{eq:app:P_pp_d}
\end{equation} 
and
\begin{align} 
\label{eq:Ldwave}
 L^{\mathrm{d},\Lambda}_{\bs{Q},\Omega}(\nu_1,\nu_3) =
 -\mathcal{D}^{\Lambda}_{\bs{Q},\Omega}(\nu_1,\nu_3)
 + \frac{1}{2}\int_{\bs{p}} \left(\cos{p_x}+\cos{p_y}\right) \Big[ 
 & \mathcal{M}^{\Lambda}_{\bs{p},\nu_3-\nu_1}(\nu_1,\Omega-\nu_1) 
 + \frac{1}{2} \mathcal{M}^{\Lambda}_{\bs{p},\Omega-\nu_1-\nu_3}(\nu_1,\Omega-\nu_1)
 \\ \nonumber
 & - \frac{1}{2} \mathcal{C}^{\Lambda}_{\bs{p},\Omega-\nu_1-\nu_3}(\nu_1,\Omega-\nu_1) \Big].
\end{align} 
Since $\mathcal{D}$ is generated exclusively by fluctuation contributions (not by the bare $U$), see Eq. (\ref{eq:Ldwave}), it is the most sensitive channel to approximations on the frequency dependence.  
Neglecting the frequency dependence of the vertex one likely overestimates $L^{\mathrm{d}}$, as already mentioned in Ref.~\onlinecite{Husemann2012}.

The flow equation for the charge channel reads
\begin{equation}
\dot{\mathcal{C}}^{\Lambda}_\bs{Q,\Omega}(\nu_1,\nu_2) = -T\sum_\nu
 L^{\mathrm{c},\Lambda}_{\bs{Q},\Omega}(\nu_1,\nu) P^{\Lambda}_{\bs{Q},\Omega}(\nu) 
 L^{\mathrm{c},\Lambda}_{\bs{Q},\Omega}(\nu,\nu_2-\Omega), 
\end{equation} 	   
with $P^{\Lambda}_{\bs{Q},\Omega}(\omega)$ as in Eq.~(\ref{eq:Pph}), and
\begin{align}  
 L^{\mathrm{c},\Lambda}_{\bs{Q},\Omega}(\nu_1,\nu_2) & =
 U - \mathcal{C}^{\Lambda}_{\bs{Q},\Omega}(\nu_1,\nu_2)
 + \int_{\bs{p}} \Big [
 - 2 \mathcal{S}^{\Lambda}_{\bs{p},\nu_1+\nu_2}(\nu_1,\nu_2-\Omega) + \mathcal{S}^{\Lambda}_{\bs{p},\nu_1+\nu_2}(\nu_1,\Omega+\nu_1)
 \nonumber \\ 
 & +  [\cos(Q_x)+\cos(Q_y)]\left( \mathcal{D}^{\Lambda}_{\bs{p},\nu_1+\nu_2}(\nu_1,\nu_2-\Omega) -\frac{1}{2} \mathcal{D}^{\Lambda}_{\bs{p},\nu_1+\nu_2}(\nu_1,\Omega+\nu_1) \right)
 \nonumber \\
 & + \frac{3}{2} \mathcal{M}^{\Lambda}_{\bs{p},\nu_2-\nu_1-\Omega}(\nu_1,\nu_2)
 + \frac{1}{2} \mathcal{C}_{\bs{p},\nu_2-\nu_1-\Omega}(\nu_1,\nu_2) \Big] .
 \label{eq:Lc}
\end{align}
The equation for the magnetic channel is reported in Eq. (\ref{eq:FlowMag}).
The form factor decomposition allows to decouple the momentum integrals, in the calculation of the $L$'s, Eqs.~(\ref{eq:Lxph}), (\ref{eq:Lswave}), (\ref{eq:Ldwave}) and (\ref{eq:Lc}), from the frequency summations in the flow equations, hence reducing the numerical effort.   

\end{widetext}

\end{appendix}

%

\end{document}